\documentclass[aps,pra,twocolumn,amsmath,amssymb,footinbib,showpacs,longbibliography,superscriptaddress
]{revtex4-2}

\usepackage[english]{babel}
\usepackage{latexsym}
\usepackage{graphics}
\usepackage{subfigure}
\usepackage{epsfig}
\usepackage{color}
\usepackage{hyperref}
\usepackage{braket} 
\usepackage[T1]{fontenc}
\usepackage[latin9]{inputenc}
\usepackage{amstext}
\usepackage{amssymb}
\usepackage{graphicx}
\usepackage{esint}
\usepackage{braket}
\usepackage{babel}
\usepackage{amsmath}
\usepackage{siunitx}

\hypersetup{
colorlinks=true,
citecolor=blue,
linkcolor=red,
urlcolor=black
}


\newcommand{\e}{\textrm{e}}
\newcommand{\ie}{i.e.}

\newcommand{\IPR}{\textrm{IPR}}
\newcommand{\Er}{E_{\textrm{r}}}
\newcommand{\Vc}{V_{\textrm{c}}}

\newcommand{\vect}[1]{{\mathbf{#1}}}
\newcommand{\rr}{\mathbf{r}}
\newcommand{\dd}{\mathrm{d}}

\begin{document}

\title{
Localization and spectral structure in two-dimensional quasicrystal potentials
}

\author{Zhaoxuan Zhu}
\affiliation{CPHT, CNRS, Ecole Polytechnique, IP Paris, F-91128 Palaiseau, France}

\author{Shengjie Yu}
\affiliation{CPHT, CNRS, Ecole Polytechnique, IP Paris, F-91128 Palaiseau, France}

\author{Dean Johnstone}
\affiliation{CPHT, CNRS, Ecole Polytechnique, IP Paris, F-91128 Palaiseau, France}

\author{Laurent Sanchez-Palencia}
\affiliation{CPHT, CNRS, Ecole Polytechnique, IP Paris, F-91128 Palaiseau, France}

\date{\today}

\begin{abstract}
Quasicrystals, a fascinating class of materials with long-range but nonperiodic order, have revolutionized our understanding of solid-state physics due to their unique properties at the crossroads of long-range-ordered and disordered systems.
Since their discovery, they continue to spark broad interest for their structural and electronic properties.
The quantum simulation of quasicrystals in synthetic quantum matter systems offers a unique playground to investigate these systems with unprecedented control parameters.
Here, we investigate the localization properties and spectral structure of quantum particles in 2D quasicrystalline optical potentials.
While states are generally localized at low energy and extended at high energy, we find alternating localized and critical states at intermediate energies.
Moreover, we identify a complex succession of gaps in the energy spectrum.
We show that the most prominent gap arises from strongly localized ring states, with the gap width determined by the energy splitting between states with different quantized winding numbers. In addition, we find that these gaps are stable for quasicrystals with different rotational symmetries and potential depths, provided that other localized states do not enter the gap generated by the ring states.
Our findings shed light on the unique properties of quantum quasicrystals and have implications for their many-body counterparts.
\end{abstract}

\maketitle

\section{Introduction}
The discovery of quasicrystals in the early 1980s~\cite{shechtman1984} marked a paradigm shift in crystallography and solid-state physics, prompting investigations into their structural and electronic properties~\cite{levine1984,senechal1995,stadnik1998,steuer2004,suck2013,steurer2018}.
Quasicrystals are usually synthetized in the laboratory after fast solidification of certain alloys~\cite{shechtman1984,shechtman1985}, but have also been observed at their natural state in meterorites~\cite{bindi2012,bindi2015} and residues of nuclear blasts~\cite{bindi2021}.
Unlike conventional crystal solids, quasicrystals lack translational invariance, but retain long-range order, hence challenging the traditional notions of crystalline order~\cite{walter2009,schmid2013}.
According to the crystallographic restriction theorem, periodic crystals can only possess two-, three-, four- or six-fold rotational symmetries \cite{Landau1958,kittel1996}. Quasicrystals, however, show discrete rotation symmetries of orders $n=5,7,8,9,...$, hence incompatible with periodic long-range order~\cite{senechal1990}. This is evidenced by Bragg spectroscopy patterns with clear diffraction peaks arranged in accordance to the forbidden rotational symmetries \cite{mackay1982,mackay1981}. The key feature of quasicrystals is then their quasi-periodicity, which amounts to the fact that finite patterns can be approximately reproduced at arbitrary long distances, but usually not exactly and not periodically.
This crystallographic definition is now pivotal in identifying and classifying quasicrystals.
Due to their structure, quasicrystals can also possess distinct electronic properties that set them apart from standard materials.
The electronic behavior in quasicrystals is governed by complex wave interference patterns, which can, for instance, give rise to phason quasiparticles \cite{edagawa2000,socolar1986,bancel1989}, exotic transport properties \cite{kamiya2018,poon1992,mayou1993,ahn2018,yao2018b} and an intricate energy spectrum \cite{cubitt2015,Fujiwara1993,Odagaki1986}.
Understanding these distinctive electronic properties is vital for unravelling the underlying mechanisms governing quasicrystals and their potential applications.

Recent developments in quantum simulation offers a unique playground to study the physics of complex quantum systems in controlled environments, with the promise of shedding light on their fundamental properties~\cite{buluta2009,NaturePhysicsInsight2012cirac, *NaturePhysicsInsight2012bloch,*NaturePhysicsInsight2012blatt,*NaturePhysicsInsight2012aspuru-guzik,*NaturePhysicsInsight2012houck,gross2017,lsp2018,*tarruell2018,*aidelsburger2018,*lebreuilly2018,*LeHur2018,*bell2018,*alet2018}.
In recent years, considerable progress has been made in the quantum simulation of quasicrystals, leveraging various experimental platforms, including
photonic crystals~\cite{chan1998,lahini2009,freedman2006,vardeny2013},
quantum fluids of light~\cite{tanese2014,barboux2017,goblot2020},
and ultracold quantum gases~\cite{lsp2005,viebahn2019}.
Notably, the use of ultracold atoms as quantum simulators has gained prominence, owing to the unprecedented control over experimental parameters~\cite{lewenstein2007,bloch2008,chin2010,lsp2010,modugno2010,mace2016}.
While one-dimensional~(1D) quasiperiodic systems have been extensively studied in the last two decades~\cite{roth2003,
damski2003,
roati2008,
fallani2007,
roscilde2008,
roux2008,
gadway2011,
derrico2014,
gori2016,
an2018,
an2021,
yao2019,
yao2020,
lellouch2014,
iyer2013,
schreiber2015,
khemani2017,
kohlert2019,
liu2022}, much attention is now devoted to quasicrystals in dimensions higher than one~\cite{viebahn2019,szabo2020,sbroscia2020,johnstone2019,johnstone2021,sbroscia2020,gautier2021,ciardi2022,zhu2023,gottlob2023}.
Previous studies on two-dimensional~(2D) optical quasicrystals have focused on investigating the
emergence of long-range quasicrystalline order using matter-wave interferometry~\cite{lsp2005,viebahn2019},
single-particle Anderson-like localization~\cite{lsp2005,szabo2020,sbroscia2020},
as well as Bose glass physics in weakly-interacting~\cite{johnstone2019,johnstone2021,sbroscia2020,JrChiunYu2023}
and strongly-correlated Bose gases~\cite{gautier2021,ciardi2022,zhu2023}.
In this respect, the physics of single quantum particles in a quasicrystal potential plays a central role, not only for their unique properties but also for understanding the many-body problem.
On the one hand, detailed studies of a variety of 1D quasi-periodic systems have revealed exotic localization and fractal properties, which clearly distinguishes generic quasiperiodic systems from disordered ones. Even richer behavior may be expected for 2D quasicrystalline systems, and a detailed study of the specific configurations implemented on current experimental platforms is deserved.
On the other hand, understanding the single-body problem is decisive for understanding the many-body properties, especially in regards to the competition of localization and interparticle interactions, which governs quantum phase diagrams.
For instance, it has been recently shown that the emergence of a Bose glass and an insulating Mott phase in strongly-correlated Bose gases subjected to shallow quasicrystal potentials can be related to localization and spectral gaps of noninteracting systems~\cite{zhu2023}.

In this work, we investigate the localization properties and the structure of the energy spectrum for non-interacting quantum particles in two-dimensional quasicrystal potentials, as realized for ultracold-atom quantum simulators.
While quantum states are generally localized at low energies and extended at high energies, we find an alternation of localized and critical states at intermediate energies.
Finite-size scaling analysis of the inverse participation ratio unveils a power law scaling, with a non-integer fractal dimension, for systems up to sizes of several hundreds of the natural length scale (optical wavelength).
Furthermore, the energy spectrum shows a complex succession of bands and gaps, and we show that the most prominent gap is generated by localized ring states, and the gap width is controlled by the energy splitting between states with different quantized winding numbers.

We present our results as follows. A 2D optical quasicrystal with eight-fold rotational symmetry is first introduced in Sec.~\ref{sec:potential} and we provide a general overview of the localization and spectral properties in Sec.~\ref{sec:spectrum}.
The localization properties are then discussed in greater detail within Sec.~\ref{sec:localization}, including the structure of localized, critical, and extended states, and their identification from a finite-size scaling analysis of the inverse participation ratio.
Next, in Sec.~\ref{sec:gaps}, we investigate the formation of gaps in the energy spectrum and show that they originate from localized ring states with different winding numbers.
Finally, in Sec.~\ref{sec:generalization}, we extend this discussion to quasicrystal potentials with other discrete rotational symmetries, before ending with our conclusions and the implications of our results in Sec.~\ref{sec:conclusions}.

\section{Quasicrystal potential}\label{sec:potential}
\begin{figure}[t!]
        \centering
        \includegraphics[width = 0.8\columnwidth]{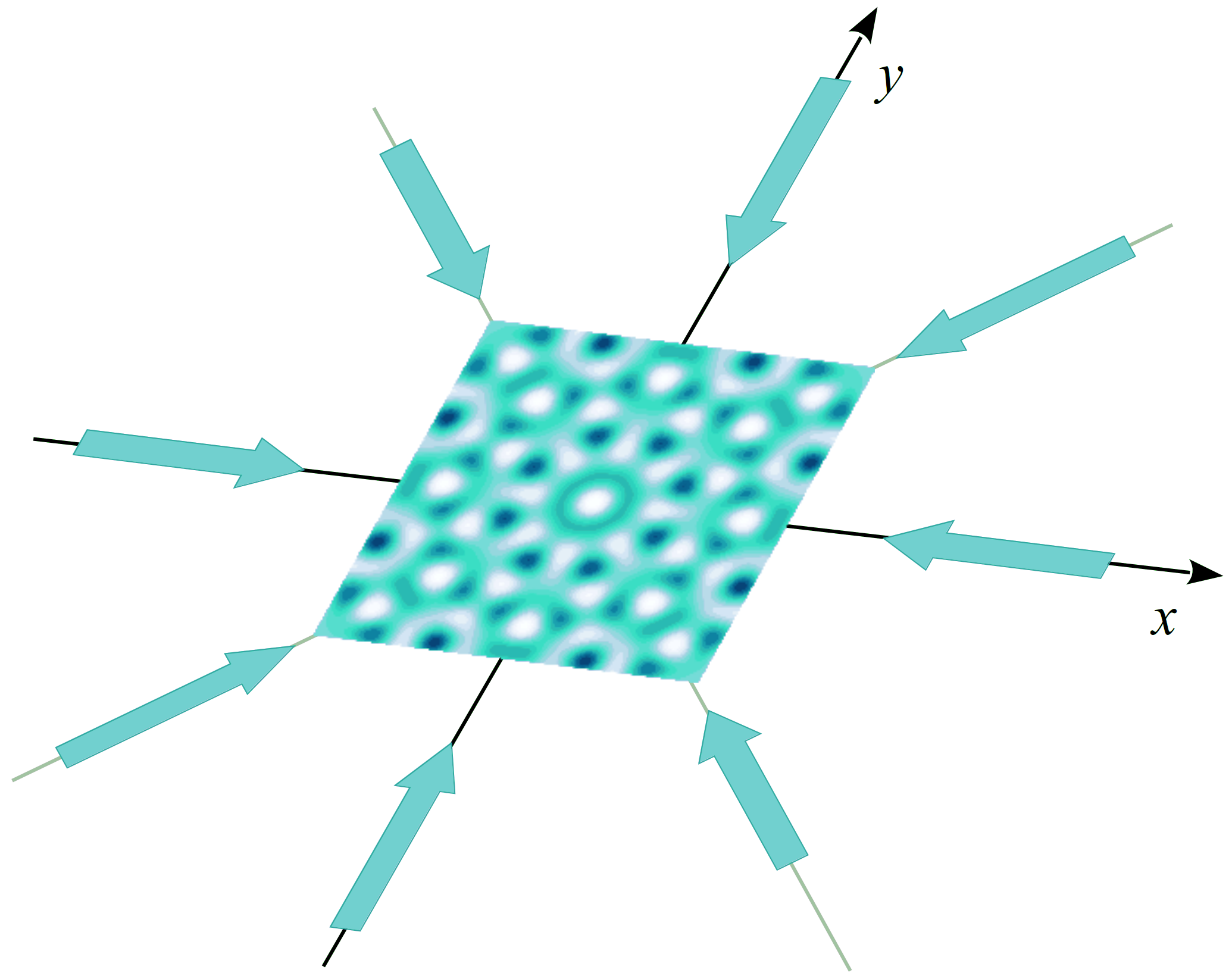}
\caption{\label{fig:potential}
Optical quasicrystal for ultracold atoms.
A 2D quasicrystal potential with eight-fold rotation symmetry is realized using four pairs of counterpropagating laser beams, making successive angles of \ang{45} each.
}
\end{figure}

The quasicrystal potential we consider, except whenever mentioned,
is the sum of four one-dimensional standing waves with amplitude $V_0$ and lattice period $a=\pi/\vert\vect{G}_k\vert$,
and successively rotated by an angle of \ang{45},
see Fig.~\ref{fig:potential}.
It can be written as
\begin{equation}\label{eq:QPpotential}
V(\rr) = V_0 \sum_{k=1}^4 \cos^2 \left(\vect{G}_k \cdot \rr +\phi_k \right),
\end{equation}
with $\vect{G}_1=\frac{\pi}{a}(1,0)$, $\vect{G}_2=\frac{\pi}{a}(\frac{1}{\sqrt{2}},\frac{1}{\sqrt{2}})$, $\vect{G}_3=\frac{\pi}{a}(0,1)$, $\vect{G}_4=\frac{\pi}{a}(-\frac{1}{\sqrt{2}},\frac{1}{\sqrt{2}})$, position vectors $\rr=(x,y)$ and phase factors $\phi_k$ of laser beam $k$.
This potential can be realized in ultracold-atom experiments using four retro-reflected laser beams with slightly shifted frequencies to suppress mutual coherence~\cite{viebahn2019,sbroscia2020,JrChiunYu2023}.
%
\begin{figure}[!t]
         \centering
         \includegraphics[width = 1\columnwidth]{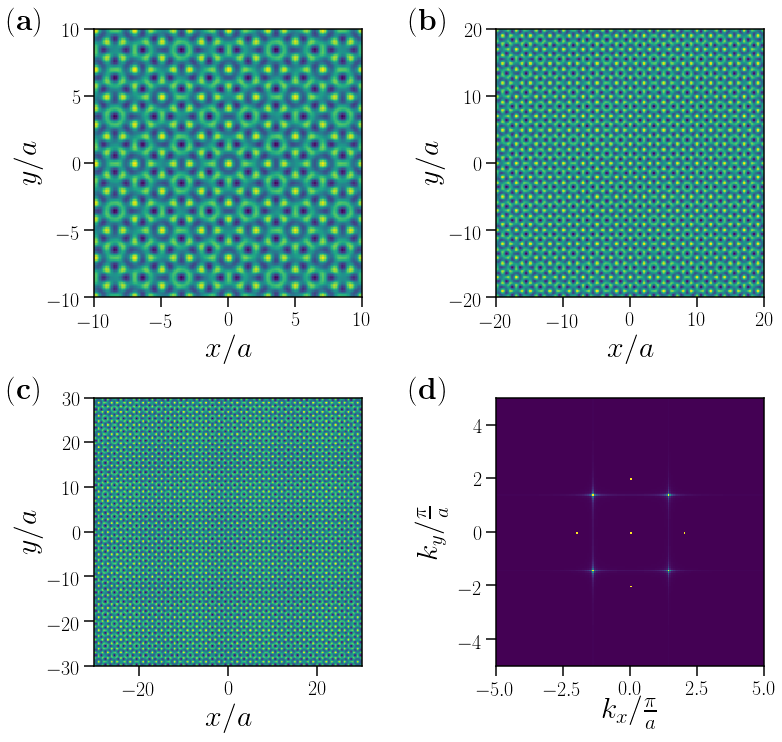}
\caption{\label{fig:potential_size}
Eight-fold quasicrystal potential.
(a)-(c)~Real-space potential, Eq.~(\ref{eq:QPpotential}) with $\phi_k=0$ for all $k$,
plotted at various scales,
for the linear system sizes
(a)~$L=20a$,
(b)~$L=40a$,
and (c)~$L=60a$,
all centered at $\rr=0$.
(d)~Fourier transform of the potental.
The color scale represents the value of the potential or its Fourier transform from low values (dark colors) to high values (light colors).
}
\end{figure}
Generalizations to quasicrystal potentials with higher rotational symmetries are discussed in Sec.~\ref{sec:generalization}.

The eight-fold quasicrystal potential with all $\phi_k=0$ is shown at different length scales on Fig.~\ref{fig:potential_size}(a), (b) and (c), which shows the quasi-repetition of short-range structures. Here, we can observe the underlying eight-fold rotational symmetry, which is incompatible with periodic order.
This is better seen in Fig.~\ref{fig:potential_size}(d), which shows the Fourier transform of the 2D potential, Eq.~(\ref{fig:potential}).
The discreteness of the Fourier pattern is a characteristic of long-range order, while the eight spots regularly arranged on a circle of radius $\lvert k \rvert =2\pi/a$ directly reflects the eight-fold discrete rotational symmetry and absence of periodic order.
For $\phi_k=0$, the origin of the 2D system at $\rr=0$ is a rotational symmetry center.
The spot at $k_x=k_y=0$ is due to the finite average value of the potential, $\int \frac{d\rr}{L^2}V(\rr)=2V_0$, where $L$ is the linear system size.
It may be cancelled out by shifting the potential by $-2V_0$.

To avoid exact rotation symmetry around the center at $(x,y)=(0,0)$, for the majority of our results, we consider an off-centered square area, which is more generic.
For instance, it lifts exact degeneracies of strongly-localized states around potential minima, that are, however, very far apart from each other.
This facilitates the discrimination of localized and extended states.
In practice, we shift the center from $(x,y)=(0,0)$ to $(x_0,y_0)=(-13543a,196419a)$, which is far beyond the system borders we consider.
This is equivalent to phase shifts of the laser beams with
$\phi_1=0$, $\phi_2\simeq 0.8597\pi$, $\phi_3=\pi$, and $\phi_4\simeq1.5540\pi$.
Note that the direction of the symmetry center to the system center is $\theta=\arctan(y_0/x_0)\simeq \ang{-86.06}$. It is away from any special directions associated with the discrete rotation symmetry, which are multiples of $\ang{22.5}$.

\section{Single particle spectrum}\label{sec:spectrum}
We now consider massive quantum particles in the quasicrystal potential of Eq.~(\ref{eq:QPpotential}).
The single-particle Hamiltonian is
\begin{equation}\label{eq:Hamiltonian}
H = \frac{\mathbf{p}^2}{2M}+V(\rr),
\end{equation}
where $\mathbf{p}=-i\hbar\boldmath{\nabla}$ is the 2D momentum operator and $M$ is the particle mass.
The eigenstates of this Hamiltonian are obtained using exact numerical diagonalization.
In practice, the diagonalizations are performed in square areas of linear sizes $L$.
We use the lattice constant $a$ as the length unit and the recoil energy $\Er={\pi^2\hbar^2}/{2Ma^2}$ as the energy unit.
We use the spatial discretization $dx=0.1a$ and the typical system size is $L=60a$, except whenever mentioned. 
Due to the finite discretization, only the lowest energy eigenstates are retained.
They correspond to those whose variation scale $\lambda$ significantly exceeds the discretization $dx$.
For an eigenstate with energy $E$, the typical de Broglie wavelength is $\lambda=\frac{2\pi\hbar}{\sqrt{2ME}}$, which yields the restriction on the eigenenergy as $E/\Er \ll \frac{4}{(dx/a)^2}=400$. In practice, we consider potential amplitudes up to $V_0=10\Er$ and we keep the eigenstates up to energy $E=8V_0$.

 \begin{figure}[t!]
         \centering
         \includegraphics[width = 0.99\columnwidth]{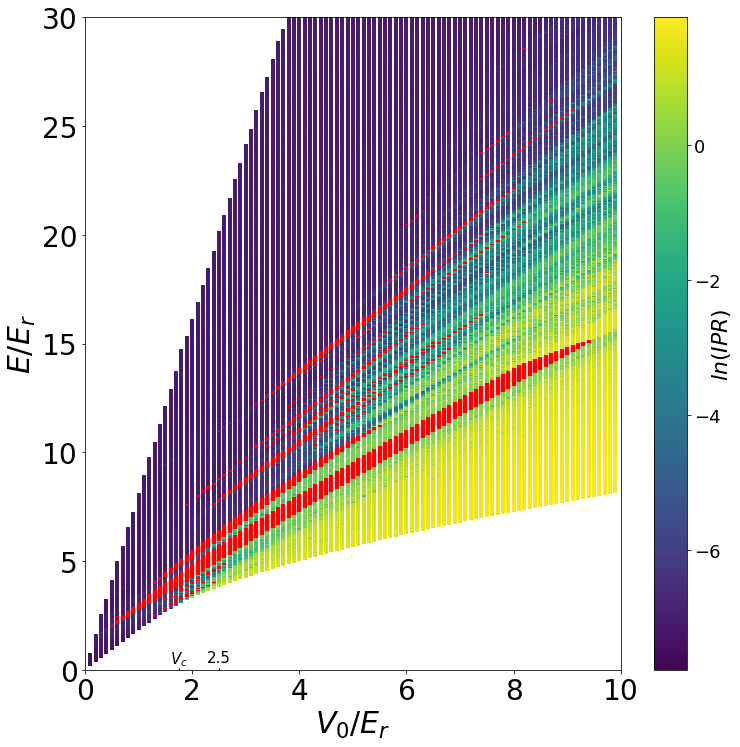}
\caption{\label{fig:ratiomin=50} 
Energy spectrum of the quasicrystal potential for various $V_0$. The system size is $L=60a$, centered at $(x_0,y_0)=(-13543a, 196419a)$.
We keep the eigenstates with energy up to $8V_0$ for potential amplitude $V_0\lesssim 4\Er$ and up to $30\Er$ for larger potential $V_0$.
The color scale shows the value of $\ln(\IPR)$ for each eigenstate, while the gaps are colored red.
}
 \end{figure}

In the numerics, we must impose boundary conditions. Owing to the nonperiodicity of the potential, any choice would distort the wavefunctions in the vicinity of the edges, and some care should be taken over the states obtained in finite squares.
We should indeed keep the states that faithfully represent the spectrum in the thermodynamic limit, and discard the states that are created by boundary effects.
Here, we choose periodic boundary conditions, where non-physical edge states may appear in the vicinity of the system boundaries.
We have checked that such edge states disappear from their initial location when we repeat the diagonalization in a larger system; while new edge states appear near the new boundaries, confirming that they are indeed created by the boundaries.
To get a spectrum representative of the thermodynamic limit, these edges states are thus excluded as follows:
We first reduce each wavefunction to a binary map.
If the wavefunction magnitude at a certain position is larger than $10\%$ of the maximum of the wavefunction magnitude, it maps to $1$; Otherwise it maps to $0$.
Then, at each point with value $1$, we compute the quantity $Z=1-{2d}/{L}$, where $d$ the distance to the nearest edge, as well as, for each eigenstate, its average $\overline Z$.
The criterion used in this work to only keep bulk states is the combination of the following two:
(i)~$\overline Z < 0.9$ and
(ii)~the position of the wavefunction maximum is at a distance larger than $3a$ from the nearest edge of the system.
If an eigenstate fulfills both conditions, it is identified as a legitimate bulk state; Otherwise, it is identified as an edge state and is excluded.

Figure~\ref{fig:ratiomin=50} shows the single-particle spectrum of the Hamiltonian versus the quasicrystal amplitude $V_0$ and the eigenenergy $E$, up to $8\Er$ (the color scale represents the $\IPR$ of each eigenstate, see Sec.~\ref{sec:localization}).
The spectrum has a rich structure and shows a series of energy gaps (highlighted in red), each in specific ranges of the quasicrystal amplitude.
The largest gap has been identified earlier~\cite{gottlob2023,zhu2023}, while the smallest are more elusive.
To locate these gaps systematically, we apply the following method for each spectrum corresponding to different values of $V_0$:
For each set of successive $500$ eigenstates, we calculate all the eigenenergy differences between neighbouring bulk states and
then take the average value of all these eigenenergy differences. If any eigenenergy difference is larger than $50$ times that mean value, it is identified as a gap.
Otherwise, it is considered to be in an energy band.
We have checked that the gaps thus identified for a system of linear size $L=60a$
agree with another approach where the energy resolution is fixed, and are
stable against increasing system sizes, see Appendix~\ref{sec:appendix.gaps}.

\section{Localization properties}\label{sec:localization}
Quasiperiodic systems are known to exhibit localization of eigenstates~\cite{aubry1980,anderson1958,biddle2009,saul1988,hiramoto1989,yao2019}.
To study these properties, we compute, for each single-particle eigenstate, the inverse participation ratio (IPR),
\begin{equation}
    P^{-1}=\frac{\int d\rr \lvert \psi(\rr) \rvert^4 }{(\int d\rr \lvert \psi(\rr) \rvert^2)^2},
\end{equation}
where $\psi(\rr)$ is the 2D eigenstate wavefunction and $P^{-1}$ represents the IPR.
Generally, in sufficiently large systems, states with large $\IPR$ are localized while states with small $\IPR$ are delocalized, which may be either extended or critical.
Figure~\ref{fig:ratiomin=50} shows, for every eigenstate, $\ln(\IPR)$ versus $V_0$ and $E$ in color scale. For small quasicrystal amplitude $V_0$, all the eigenstates appear delocalized.
This is consitent with the critical potential for localization $\Vc \simeq 1.76\Er$, found in previous works~\cite{szabo2020,gautier2021}.
As $V_0$ increases, the low energy eigenstates tend to localize since long-range coherence is suppressed as the quasicrystal becomes deeper, while higher energy eigenstates remain delocalized. 
In the intermediate energy range, the localization behaviour is richer.
We find that eigenstates with large and small $\IPR$ coexist, and in particular, the $\IPR$ is nonmonotonous against energy.

\subsection{Finite-size scaling analysis}
To study localization in more detail, the bare value of the $\IPR$ is insufficient and we consider a more rigorous characterization of localization.
It is provided by the scaling of $\IPR$ with the system size $L$,
$\IPR \sim L^{\gamma}$.
In 2D, localized states are characterized by $\gamma=0$ and extended states by $\gamma=-2$.
Any intermediate value of $\gamma$ between $0$ and $-2$ then identifies a critical state \cite{saul1988,hiramoto1989}.
In the following, we focus on the case $V_0=2.5\Er$, which is larger than the critical potential $\Vc\simeq 1.76\Er$, but similar results are found for other amplitudes of the quasicrystal potential.

Studying the finite-size scaling properties of the $\IPR$ requires comparing eigenstates for different sizes. However, diagonalization in systems with different sizes gives different total numbers of eigenstates, as the density of states generally scales with the system area.
Hence, given an eigenstate for a certain system size, there is always some arbitrariness on picking up the corresponding eigenstate for another system size to be compared with.
To overcome this issue, we compute the averaged $\IPR$ of all eigenstates in a narrow energy window, $\overline{\IPR}$. The width of the energy window is chosen to be $\delta E = 0.01\Er$, narrow enough such that the eigenstates in the energy window have similar localization properties, but large enough to have enough states to average.
Then, the scaling exponent $\gamma$ considered below refers to this averaged $\overline{\IPR}$, and characterizes the localization properties in the corresponding energy window. 

\begin{figure}[t!]
         \centering
         \includegraphics[width = 0.99\columnwidth]{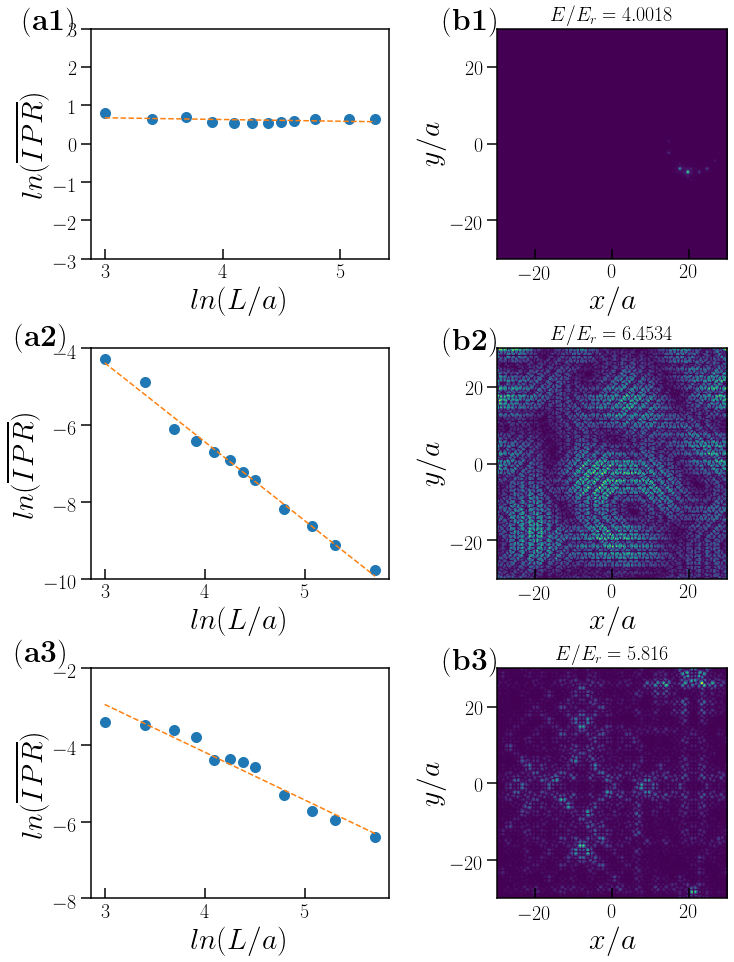}
\caption{\label{fig:scaling_single} 
Localization properties of  some eigenstates.
The left column, (a), shows $\ln(\overline{\IPR})$ versus $\ln(L/a)$ for all eigenstates in various energy windows. 
Blue disks are data points while orange dash lines show linear fits.
The quasicrystal amplitude is $V_0=2.5\Er$ and the system sizes for linear regressions range from $L=20a$ to $L=300a$.
(a1)~The energy window is $E/\Er \in [4.00,4.01]$ and the linear fit yields $\gamma=-0.04\pm0.03$.
(a2)~The energy window is $E/\Er \in [6.45,6.46]$ and the linear fit yields $\gamma=-2.06\pm0.06$.
(a3)~The energy window is $E/\Er \in [5.81,5.82]$ and the linear fit yields $\gamma=-1.25 \pm 0.09$.
The right column, (b), show examples of eigenstates in the corresponding energy windows, with energies indicated on the top of each panel.
The wavefunctions plotted here are computed for the system size $L=60a$.
}
 \end{figure}

For instance, Fig.~\ref{fig:scaling_single} shows the scaling of the $\IPR$ versus the system size from $L=20a$ to $L=300a$, for three different energy windows.
 In all cases, it shows a clear power-law scaling, $IPR\sim L^\gamma$, and the exponent $\gamma$ is found by a linear fit of
$\ln(\overline{\IPR})$ versus $\ln(L/a)$.
Figure~\ref{fig:scaling_single}(a1) corresponds to energy $E \simeq 4\Er$.
The quantity $\ln(\overline{\IPR})$ fluctuates with the system size but shows no clear increasing or decreasing tendency. 
Linear regression of the data points yields the slope $\gamma=-0.04\pm0.03$.
The small value of $\lvert \gamma \rvert$ indicates that the states in this energy window are localized.
An example of the wavefunction of an eigenstate in this energy window is plotted in Fig.~\ref{fig:scaling_single}(b1). 
The state is localized in a few local potential wells, consistently with strong localization.
Figure~\ref{fig:scaling_single}(a2) corresponds to energy $E \simeq 6.45\Er$.
In this case, linear regression yields the slope $\gamma=-2.06\pm0.06$, corresponding to an extended state.
The wavefunction of an eigenstate in this energy window, shown in Fig.~\ref{fig:scaling_single}(b2), consistently covers the full system area, although not homogeneously.
Finally, Fig.~\ref{fig:scaling_single}(a3) corresponds to energy $E \simeq 5.81\Er$. 
It also shows a clear linear behaviour in log-log scale and the slope is found to be $\gamma=-1.25 \pm 0.09$, significantly far from both $0$ or $-2$.
In this energy window, the states are thus neither localized nor extended, \ie~they are critical. A typical eigenstate is plotted in Fig.~\ref{fig:scaling_single}(b3).
Unlike localized states, these critical states extend over the full system
but, compared to extend states, they only cover a limited proportion of the area. Note, in general, $\gamma$ may exhibit quite large standard deviations, particularly when one considers domains where critical states appear. In our calculations, we consider the average value and standard deviation from the linear fit, and do not take into account the spread of individual IPR values. As we discuss in Appendix~\ref{Appendix:QualityFitIPR}, the distribution of the IPR for localized and extended states is rather narrow, and does not contribute significant errors. In contrast, for critical states, we tend to observe a broader distribution of coefficients $\gamma$ even in rather narrow energy windows, which, however, remains clearly distinct from either localized or extended states.
This behaviour is strongly indicative of a critical regime, despite the significant spread of IPR values (and hence larger error in $\gamma$).
The value of $\gamma$ should therefore be read as a qualitative figure within critical domains, rather than a quantitative one.

\begin{figure}[t!]
         \centering
         \includegraphics[width = 0.99\columnwidth]{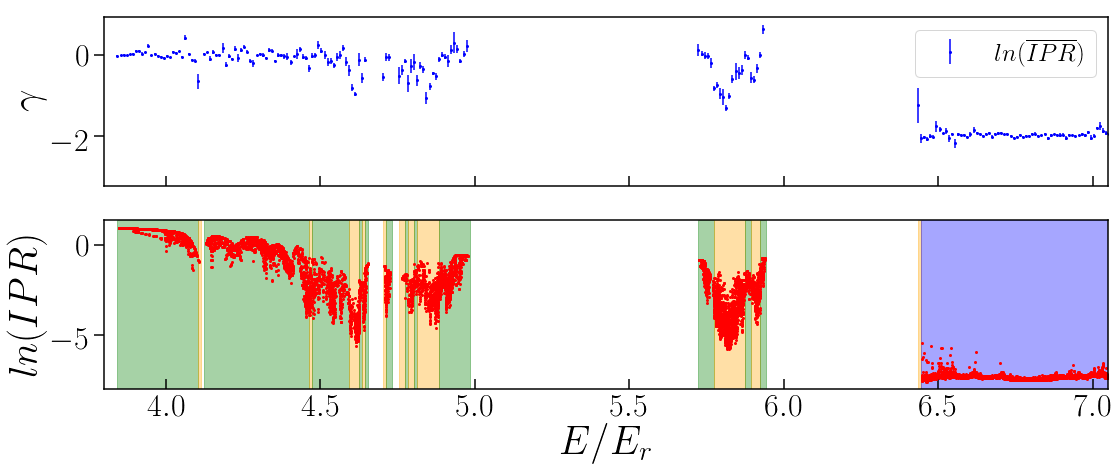}
\caption{\label{fig:scaling_full}
Localization spectrum in the quasicrystal potential.
Upper row:~Scaling exponent $\gamma$ of the $\IPR \sim L^\gamma$ for system sizes ranging from $L=20a$ to $L=100a$,
except in some narrow energy windows where data for larger sizes, $L=120a,160a,200a,300a$, are also calculated.
Lower row:~Classification of different kinds of states. The values of $\ln(\IPR)$ for all eigenstates with system size $L=80a$ is plotted versus their eigenenergies. Localization properties are shown as background colors: green for localized states, blue for extended states, and orange for critical states.
The potential amplitude is $V_0=2.5\Er$.
}
 \end{figure}

Systematic finite size scaling and linear fits are performed over the full spectrum, with results shown in Fig.~\ref{fig:scaling_full}.
For details, see Appendix~\ref{Appendix:QualityFitIPR}.
Generally, the system sizes range from $L=20a$ to $L=100a$. For some energy windows, notably where critical states appear, we use larger sizes, up to $L=300a$, so as to check that the scaling behaviour of the $\IPR$ persists for larger system sizes.
The exponent $\gamma$ found from linear fits of $\ln(\overline{\IPR})$ versus $\ln(L/a)$ is shown in the upper row of Fig.~\ref{fig:scaling_full}. 
The results show that the lowest energy states are localized, with $\gamma \simeq 0$.
For energy $E \gtrsim 6.45\Er$, \ie~above the second large energy gap, all states are extended, with $\gamma \simeq -2$.
In the intermediate energy range, there are critical states whose scaling exponent $\gamma$ is neither $0$ nor $-2$ but clearly in between.
Taking into account the uncertainty of the fitted exponents $\gamma$,
energy ranges with different kinds of localization properties can be identified
as follows, with results shown in the lower row of Fig.~\ref{fig:scaling_full}:
Localized states for $\gamma > -0.25$ (green),
extended states for $\gamma < -1.75$ (blue),
and critical states for $-1.75<\gamma<-0.25$ (yellow).
Also shown is $\ln(\IPR)$ for each eigenstate calculated for $L=80a$ (red dots).
The behaviour of the DOS shows that that there is a significant number of states of each kind in the spectrum. 
The behaviour of $\ln(\IPR)$ is rather smooth, up to significant fluctuations.
Interestingly, we find that critical and localized states can coexist at intermediate energies, with no clear mobility edge, or separation between localized and critical domains.
This is reminiscent of ``anomalous mobility edges'' separating bands of localized states and bands of critical states as found in other quasiperiodic models~\cite{liu2022}.

General properties of localized, extended, and critical states are further discussed in the following subsections.

 \begin{figure}[t!]
         \centering
         \includegraphics[width = 0.99\columnwidth]{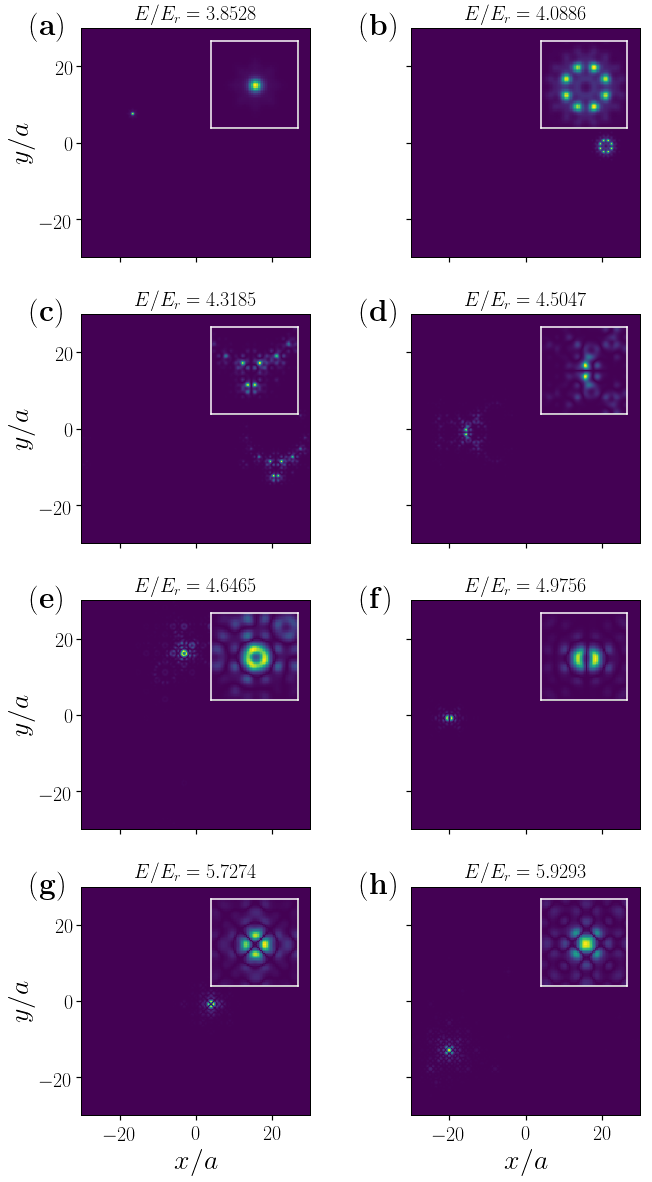}
\caption{\label{fig:wavefun_localized} 
Wavefunctions of typical localized states in various energy ranges.
The eigenenergy, indicated on the top of each panel, increases in the reading order, from~(a) to (h).
The quasicrystal potential amplitude is $V_0=2.5\Er$ and the system size is $L=60a$.
The insets show magnifications of the corresponding panel in the vicinity of the localization center.
}
 \end{figure}

\subsection{Localized states}
Localized states generally appear in the low energy range, as well as at some energy band edges, near by energy gaps, see Fig.~\ref{fig:scaling_full}. Typical localized states in different energy ranges are plotted on Fig.~\ref{fig:wavefun_localized}, with eigenenergy increasing in the reader order, from panel~(a) to panel~(h). The insets show magnification of the main panel in the vicinity of the localization center.
The states with lowest energies are strongly localized in a single local potential well, see Fig.~\ref{fig:wavefun_localized}(a).
As the energy increases, the states start to cover a few potential wells.
Some states are localized in regions where the local potential is almost eight-fold rotational symmetric, and the eigenstates form rings composed of eight almost equivalent spots, see Fig.~\ref{fig:wavefun_localized}(b).
Similar ring states exist with 16 spots (not shown).
Other localized states cover a small cluster of different potential wells, such that the tunnelling between them is large enough to compensate the eigenenergy differences between the local potential wells, see Figs.~\ref{fig:wavefun_localized}(c) and (d).
As the energy further increases, states composed of one or many small rings begin to appear. Figure~\ref{fig:wavefun_localized}(e) shows such a localized state with one small ring, with eigenenergy slightly below the energy gap at $E \simeq 4.65\Er$. The small ring actually corresponds to a set of several local shallow potential wells that are very near by each other, and the wavefunction on these wells merge into an almost homogeneous circle.
Figures~\ref{fig:wavefun_localized}(f) and (g) are localized states with energies, respectively, right below and right above the first large energy gap in, approximately, $E/\Er \in [5,5.7]$. They are also states located on small rings of nearby potential wells, as for Fig.~\ref{fig:wavefun_localized}(e).
However, unlike in Fig.~\ref{fig:wavefun_localized}(e), these small-ring wavefunctions have, respectively, one or two node lines due to different phase winding numbers. These states play special roles in the spectral structure and will be further discussed in Sec.~\ref{sec:gaps}.
Localized states with relatively high energies do not all have similar ring structures.
Some of these states are localized in one or multiple potential wells, as in Fig.~\ref{fig:wavefun_localized}(h).

\subsection{Extended states}
 \begin{figure}[t!]
         \centering
         \includegraphics[width = 0.99\columnwidth]{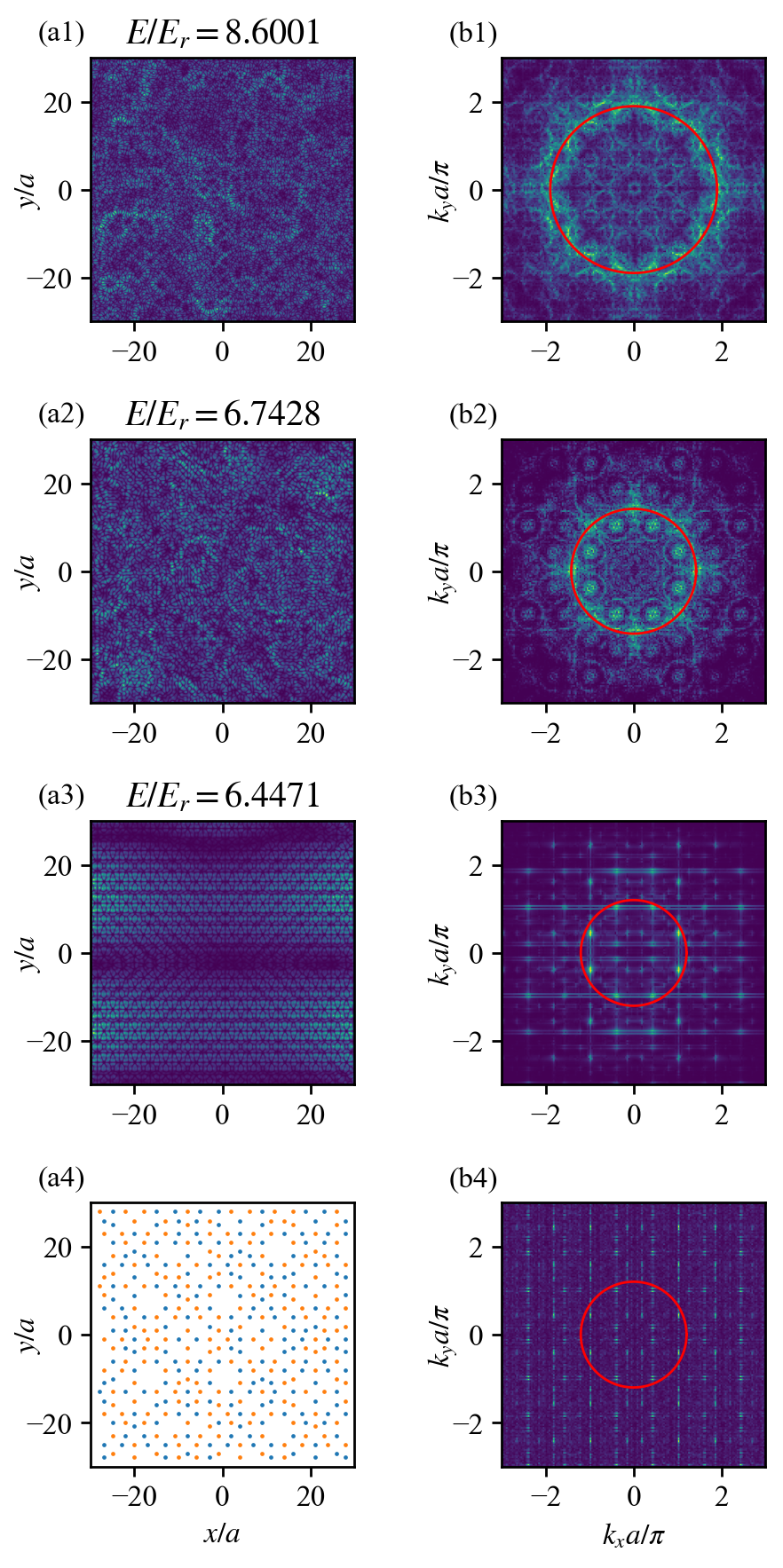}
\caption{\label{fig:wavefun_extended}
Wavefunctions of extended states, with energies given on top of each panel, $V_0=2.5\Er$, and $L=60a$.
We show the (a1)-(a3) real-space density profiles and corresponding (b1)-(b3) Fourier space momentum distributions.
The red circle indicates the momentum $k_E$ given by a perturbation theory.
Panels (a4) and (b4) show the reconstructed components of (a3)-(b3), Eq.~(\ref{eq:ReconstExtStatebis}).
The spots in (a4) show the positions of the local maxima $\mathbf{r}_j$, with color indicating the phase (blue : $\phi_j=0$, orange: $\phi_j=\pi$).
Panel~(b4) shows the square modulus of the Fourier transform of Eq.~(\ref{eq:ReconstExtStatebis}).
The positions of spots in (b4) match well with those in (b3).}
 \end{figure}
States with high enough eigenenergies are all extended, see Fig.~\ref{fig:scaling_full}.
For a potential amplitude $V_0=2.5\Er$, extended states appear after the second large gap, \ie~for energy $E \gtrsim 6.45\Er$.
Typical extended states are plotted on Fig.~\ref{fig:wavefun_extended}.
They cover the full system area quasi-homogeneously, even though some dark node regions due to large scale modulation of the wavefunction may appear, as for instance in Fig.~\ref{fig:wavefun_extended}(a3). 
Extended wavefunctions contain many spots separated by node lines. 
For low-enough energy, each island has a size comparable with the lattice constant $a$, and forms a rather ordered pattern, see Fig.~\ref{fig:wavefun_extended}(a3).
For states with higher energy, the wavefunction variations become stronger and the typical size of the islands gets smaller.
It goes down to the de Broglie wavelength, $\lambda={2\pi\hbar}/{\sqrt{2mE}}$, of the quasi-plane waves rather than the potential profile.
Correspondingly, the states show a complex, quite disordered, pattern determined by multiple scattering on the quasicrystal potential, see Figs.~\ref{fig:wavefun_extended}(a1) and (a2). 

While the IPR alone does not seem to characterise the crossover from "ordered" extended states to "disordered" ones,
a better understanding can be gained by looking at the density profiles in momentum space, as shown in Figs.~\ref{fig:wavefun_extended}(b1), (b2), and (b3). Consider first the highest energy states [Figs.~\ref{fig:wavefun_extended}(b1) and (b2)].
The momentum distribution of such states is concentrated around a marked circle, with a smaller spreading when the energy increases.
This structure may be understood using simple perturbation theory: Extended states with high-enough energy $E$ are constructed from plane waves with momentum $k_E$ such that $E=\hbar^2 k_E^2 /2M + \langle V \rangle$, where $\langle V \rangle \simeq 2V_0$ is the potential-energy contribution for purely plane waves. We consistently find that the circle with radius $k_E$ [shown in red in Figs.~\ref{fig:wavefun_extended}(b1), (b2), and (b3)] coincides with the dominant momentum components of the extended wavefunction. The quasicrystal potential weakly couples many plane waves with a modulus of the momentum nearly equal to $k_E$ in almost all directions and within a structure consistent with the quasicrystalline eightfold rotational symmetry.

The formation of the momentum profile of extended states at a smaller energy is more subtle. In this case, the extended wavefunctions are dominated by the quasicrystal potential and we now have a quasiperiodic, lattice-like structure [Fig.~\ref{fig:wavefun_extended}(b3)]. The origin of this structure can be deduced as follows. First, we note that the extended state in real space is made of hybridized ring states with 3 nodes lines.
The rings are located around some potential maxima, which forms a discrete lattice in real space, shown in Fig.~\ref{fig:wavefun_extended}(a4).
The wavefunction can thus be approximated as
\begin{equation}\label{eq:ReconstExtState}
\psi(\mathbf{r}) \simeq \int d\mathbf{r}' \psi_0(\mathbf{r}) f(\mathbf{r} - \mathbf{r}')
\end{equation}
and
\begin{equation}\label{eq:ReconstExtStatebis}
f(\mathbf{r}) = \sum_j e^{i\phi_j} \delta(\mathbf{r}-\mathbf{r}_j),
\end{equation}
where $\psi_0(\mathbf{r})$ is a sample of the ring states with 3 node lines, $\mathbf{r}_j$ is the position of the $j$th potential maxima and $\phi_j$ is a phase factor of either $0$ or $\pi$, which gives the correct alignment of the local structures in the original wavefunction.
The phase factor is encoded in the color (blue or orange) of the lattice points in Fig.~\ref{fig:wavefun_extended}(a4).
To check this interpretation, we compute the Fourier transform of $f(\mathbf{r})$ in Eq.~(\ref{eq:ReconstExtStatebis}) and plot it in Fig.~\ref{fig:wavefun_extended}(b4). 
The result reproduces well the primary features of the momentum profile found in Fig.~\ref{fig:wavefun_extended}(b3), namely the quasiperiodic, lattice-like structure, and some of the most significant spots.
This shows that the structure of the lowest-energy extended states is strongly related to the quasiperiodic nature of the potential.
When the energy $E$ is further increased, the quasiperiodic structure becomes less significant, with the progressive emergence of the plane-wave momentum circles observed in Figs.~\ref{fig:wavefun_extended}(a1) and (a2).

\subsection{Critical states}
\begin{figure}[t!]
         \centering
         \includegraphics[width = 0.95\columnwidth]{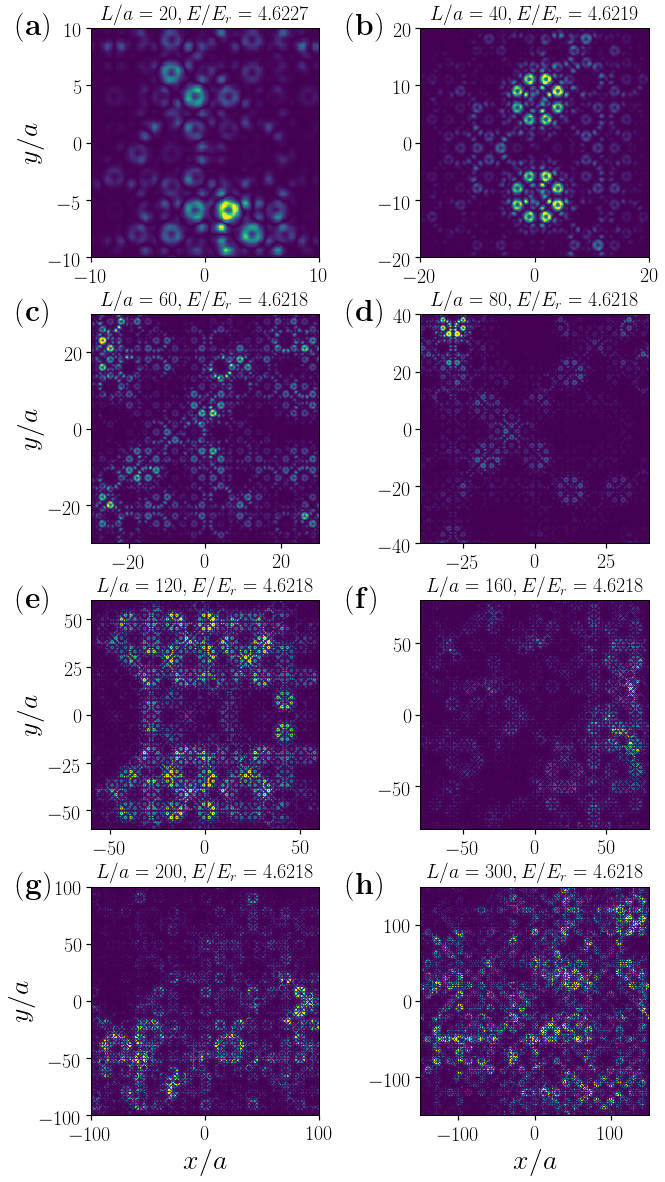}
\caption{\label{fig:wavefun_critical1} 
Wavefunctions in the critical regime at  $E \simeq 4.6218\Er$ for $V_0=2.5\Er$, shown at different scales.
The various panels show the wavefunction of the state with eigenenergy closest to $E=4.6218\Er$ for system sizes increasing in the reader order, from (a)~$L=20a$ to (h)~$L=300a$.
Note that the colors have been rescaled to increase the contrast.
}
 \end{figure}

\begin{figure}[t!]
         \centering
         \includegraphics[width = 0.95\columnwidth]{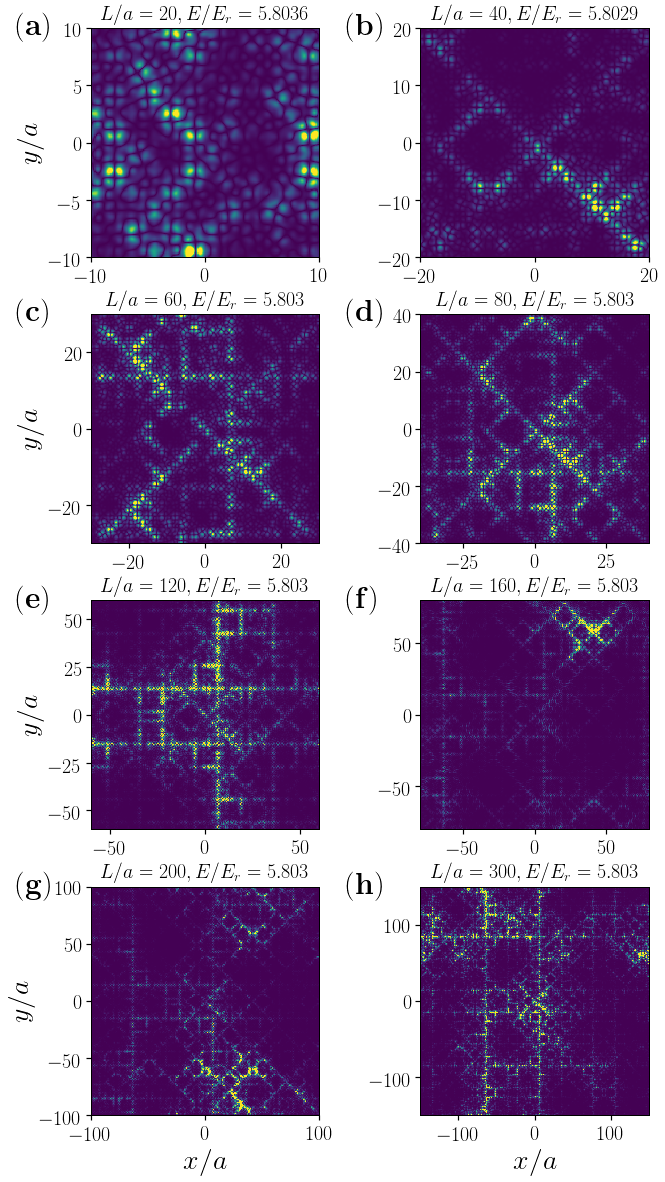}
\caption{\label{fig:wavefun_critical2} 
Wavefunctions in the critical regime at  $E \simeq 5.803\Er$ for $V_0=2.5\Er$, shown at different scales.
The various panels show the wavefunction of the state with eigenenergy closest to $E=5.803\Er$ for system sizes increasing in the reader order, from (a)~$L=20a$ to (h)~$L=300a$.
Note that the colors have been rescaled to increase the contrast.
}
 \end{figure}

Critical states generally appear in between the localized states not far from the edges of energy bands, see~Fig.~\ref{fig:scaling_full}. The critical states typically extend across the full system, with complex geometrical patterns,
separated in two main classes, illustrated in
Figs.~\ref{fig:wavefun_critical1} and \ref{fig:wavefun_critical2}.

%

On the one hand, Fig.~\ref{fig:wavefun_critical1} displays the states with energy closest to $E=4.6218\Er$ for different system sizes.
For this energy, the scaling exponent is $\gamma=-1.49\pm0.24$, clearly different from $0$ and $-2$.
As visible on Fig.~\ref{fig:wavefun_critical1}(a), these critical states are all composed of small rings, similar to localized states with no node lines in the rings as in Fig.~\ref{fig:wavefun_localized}(e).
These critical states, which contain a countless number of the small rings, have energies slightly smaller than the localized states containing only one or a few small rings, owing to hybridization, which minimizes the tunnelling energy.
Since the ring states that form the building blocks of such critical states are roughly isotropic, they do not favor any clear direction, and, on a larger scale, they group together and form larger ring structures containing eight small rings, and eventually form a ``ring of ring'', see  Fig.~\ref{fig:wavefun_critical1}(b).
On even larger scales, these rings of rings  also group together forming more complex structures, see Figs.~\ref{fig:wavefun_critical1}(c)-(h).

On the other hand, Fig.~\ref{fig:wavefun_critical2} shows an example of the other class of critical states, found at a higher energy. It displays the states with energy closest to $E=5.803\Er$ for different system sizes.
Here the scaling exponent is $\gamma=-0.87\pm0.18$, also corresponding to critical states.
However, unlike Fig.~\ref{fig:wavefun_critical1}, which contains rings at different scales,
Fig.~\ref{fig:wavefun_critical2} displays square-shaped structures.
On large enough scales, we find that these states display straight lines either along the main axes ($x$ and $y$) or along the diagonals.
In fact, these states are also built from ring states, but here with two node lines, similar to that shown on Fig.~\ref{fig:wavefun_localized}(g).
For such ring states with node lines along the diagonals, hybridization is favored along the main axes, which maximizes the overlap of wavefunctions from adjacent rings.
This explains the appearance of lines along the main axes.
Moreover, owing to the eightfold rotation symmetry of the potential, there are also ring states with two node lines, but now oriented along the main axes.
For such states, the hybridization is then favored along the diagonals, which creates the distinct lines across the diagonals.
Finally, since all the ring states with two node lines along either the main axes or along the diagonals are quasi-degenerate, square-like structures oriented in either directions also hybridize, hence forming the complex structure observed in Fig.~\ref{fig:wavefun_critical2}.
In support of this interpretation, zooms of the wavefunction show that the lines parallel to the main axes are formed of ring states with two node lines along the diagonals while the lines along the diagonals are formed of ring states with two node lines along the main axes, see~Appendix~\ref{sec:appendix.critical_square}.

For completeness, we have also inspected the critical states near $E \simeq 4.8\Er$, which now have one node line across their local centre. Some examples are shown in Appendix~\ref{sec:appendix.critical_square}.

\section{Gaps bounded by ring states with phase winding}\label{sec:gaps}
\begin{figure}[t!]
         \centering
         \includegraphics[width = 0.99\columnwidth]{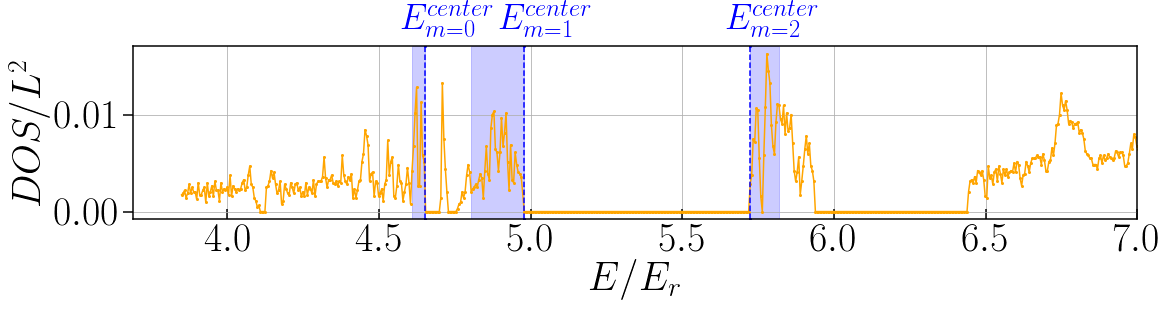}
\caption{\label{fig:spectrumgap}
Density of states per unit area versus energy for potential amplitude $V_0 = 2.5\Er$ and system size $L=80a$.
The energies of the centered small rings with winding $m=0$, $1$, and $2$ are indicated by the dashed lines. The shaded areas indicate the regions of the corresponding off-centered rings with the same winding numbers.
}
 \end{figure}

We now study the structure of the energy spectrum.
The DOS for the quasicrystal with amplitude $V_0=2.5\Er$ and a large system size, $L=80a$, is shown in Fig.~\ref{fig:spectrumgap}.
It displays two main gaps, in the energy windows $E/\Er \in [4.98,5.72]$ and $E/\Er \in [5.94,6.44]$, respectively, as well as three smaller gaps 
at $E/\Er\simeq 4.11$, $E/\Er\simeq 4.65$, and $E/\Er\simeq 4.71$.
To understand the origin of these energy gaps, we study the properties of eigenstates in the vicinity of the band edges.

\subsection{Ring states at band edges}\label{sec:gaps.rings}
States localized in small ring structures of the potential play an important role in the energy spectrum, as also observed in Ref.~\cite{gottlob2023}.
For instance, the states right before the small gap at $E \simeq 4.65\Er$ are composed of small rings with a roughly homogeneous density, similar to that shown in
Fig.~\ref{fig:wavefun_localized}(e), and spread over different locations.
The states right before the first large energy gap at $E/\Er \in [4.98,5.72]$ are similar ring states but with one node line, similar to that shown in Fig.~\ref{fig:wavefun_localized}(f).
The states right after the same large gap are also ring states but with two node lines, similar to that shown in Fig.~\ref{fig:wavefun_localized}(g). The energies of ring states with $0$, $1$ or $2$ node lines are highlighted with the shaded blue areas in Fig.~\ref{fig:spectrumgap}.

\begin{figure}[t!]
         \centering
         \includegraphics[width = 0.92\columnwidth]{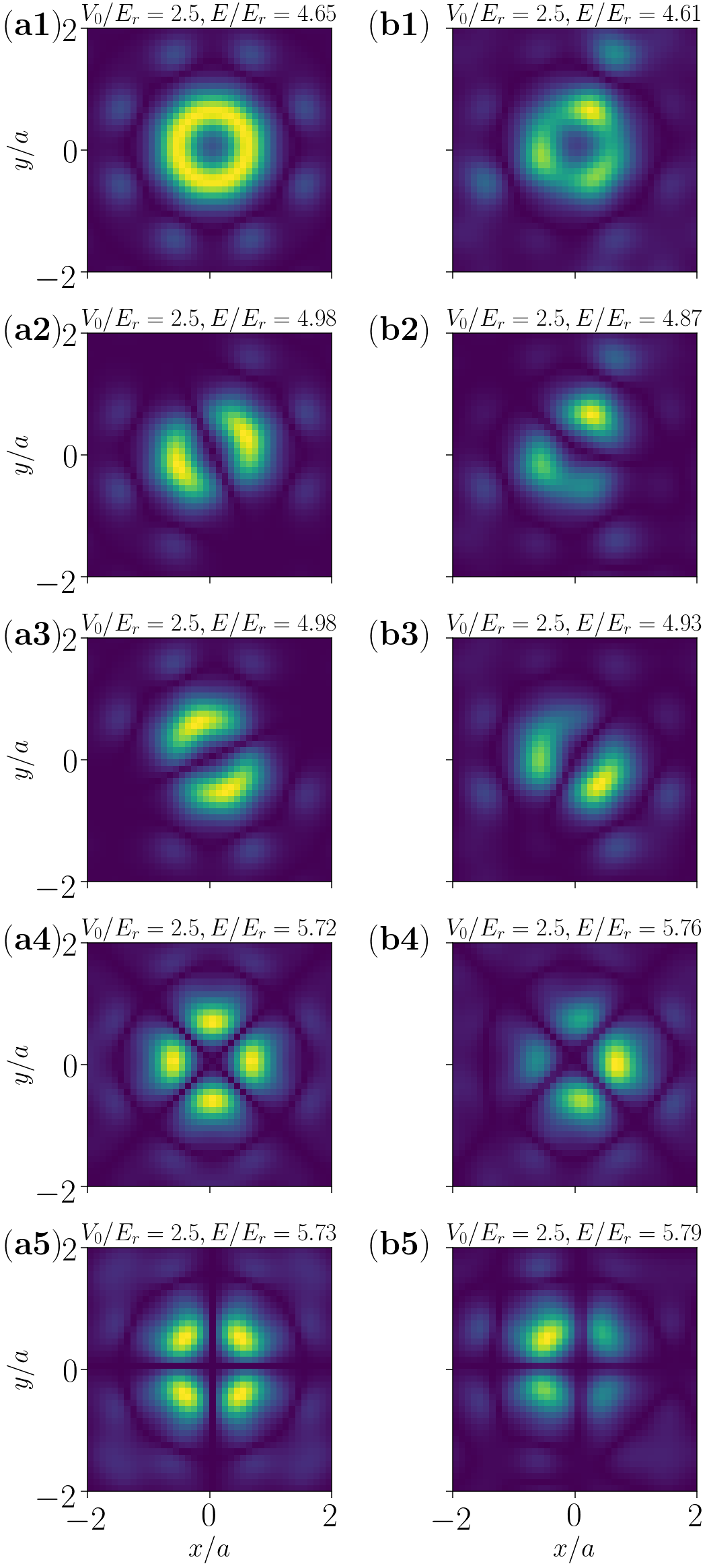}
\caption{\label{fig:ring_center}
Centered (left) and off-centered (right) ring states in the eightfold quasicrystal potential centered at $\rr_0=(0,0)$, with amplitude $V_0=2.5\Er$.
(a1)-(a5)~States at energy band edges:
(a1)~Last state before the small gap at $E/\Er\simeq 4.65$;
(a2) and (a3)~Two almost degenerate states immediately before the first large gap in $E/\Er\sim [4.98,5.72]$;
(a4) and (a5)~Two almost degenerate states immediately after the same first large gap.
They are eigenstates for the smallest ring around $\rr=0$ with winding numbers (node line numbers) $m=0, \pm 1, \pm2$.
(b1)-(b5)~Off-centered ring states around $\rr_0=(19.06a, 4.95a)$ obtained by diagonalization in a square with size $L=4a$.
}
 \end{figure}

Although we have so far excluded the symmetry center of the potential by considering off-centered systems, it is useful to re-incorporate it in the discussion of these ring states. Indeed, very regular
ring states exist for the eight-fold quasicrystal potential around the symmetry center at $\rr=0$, see Fig.~\ref{fig:ring_center}(a).
Remarkably, these centered ring states have energies that exactly limit certain bands, as indicated by the vertical, dashed blue lines in Fig.~\ref{fig:spectrumgap}.
For instance, the centered ring state with no node lines, Fig.~\ref{fig:ring_center}(a1), is the very last state of the energy band before the small gap at $E \simeq 4.65\Er$.
Then, there are two degenerate centered ring states with a single node line each, Figs.~\ref{fig:ring_center}(a2) and (a3),
which lie immediately before the first large gap, at $E \simeq 4.98\Er$.
Finally, there are another two degenerate, centered ring states with two node lines, Figs.~\ref{fig:ring_center}(a4) and (a5), immediately after the first large gap, at $E \simeq 5.72\Er$.
The other ring states with $0$, $1$ or $2$ node lines that are situated in rings away from the center of the quasicrystal are slightly distorted, see Figs.~\ref{fig:ring_center}(b1)-(b5), and their energies lie deeper in the energy bands, as indicated by the shaded areas in Fig.~\ref{fig:spectrumgap}.
As discussed below, rejection of the off-centered ring states inside the energy bands may be interpreted as a level repulsion effect.

For the sake of completeness, note that the first state after the second large gap at $E \simeq 6.44\Er$ is made of ring states with three node lines. In this case, however, many such rings are connected, hence forming a quite compact extended state. The states at the band edges of the gap at $E/\Er\simeq 4.11$ are localized states, and will be briefly discussed later in Sec.~\ref{sec:gaps.minigaps}. Finally, the states at the other band edges before $E \simeq 6.44\Er$ are also localized states, without any particular features in their patterns.

These observations indicate that the small ring states, especially the ones centered on the symmetry center of the quasicrystal potential at $\rr=0$, play special  roles in the structure of the spectrum, and we discuss them in more detail below.

\subsection{Central ring}\label{sec:gaps.centered}
We first consider the ring states around the quasicrystal symmetry center at $\rr=0$, see Fig.~\ref{fig:ring_center}(a).
They are located in a nearly isotropic annular potential well of radius  $\rho_0\simeq 0.61a$.
More precisely, the potential has eight very shallow potential wells along the ring, which may, however, be neglected.
In particular, we find that the ring-state with no node, Fig.~\ref{fig:ring_center}(a1), has a nearly isotropic density with modulations less than $10\%$.
As a result, the small ring states in Fig.~\ref{fig:ring_center}(a) may be approximated by states localized in the nearly isotropic annular potential well with almost isotropic density and $\textrm{O}(2)$ rotational symmetry around the symmetry center $\rr=0$.
In the vicinity of the annular well, the Hamiltonian and the planar angular momentum operator $\hat{L}_z$ can be diagonalized simultaneously, and the wavefunctions can be written as $\phi_m^0(\vect{r}) \simeq u_m(r)\e^{im\theta}$, where $u_m(r)$ is a real-valued function, $\theta$ is the polar angle, and $m\in\mathbb{Z}$ is the phase winding number. For sufficiently strong potential amplitude $V_0$ and low winding number $m$, the functions $u_m(r)$ are strongly confined around $r = \rho_0$ and we may neglect the $m$-dependence originating from the centrifugal term. For details, see Appendix~\ref{sec:appendix.centeredring}.
Figure~\ref{fig:ring_center}(a1) is the state with winding number $m=0$.
The states in Figs.~\ref{fig:ring_center}(a2) and (a3) correspond to winding numbers $m = \pm 1$.
Since the states $\phi_{\pm 1}^0(\vect{r})$ are strictly degenerate, any linear combination of both is also an eigenstate of the Hamiltonian. 
Numerical diagonalization returns real-valued wavefunctions,
\ie~$\psi_{+1}^{0}(\vect{r}) \simeq \sqrt{2} u(r)\cos(\theta-\theta_1)$
and $\psi_{-1}^{0}(\vect{r}) \simeq \sqrt{2} u(r)\sin(\theta-\theta_1)$,
where $\theta_1$ is some reference angle.
Consequently, the two states in Figs.~\ref{fig:ring_center}(a2) and (a3) show orthogonal node lines at the angles $\theta_1+ \pi/2$ and $\theta_1$, respectively.
The angle $\theta_1$ is determined by the small modulations of potential along the annular well and, in the numerics, by the discretization, which does not satisfy an exact eight-fold rotation symmetry.
Similarly, Figs.~\ref{fig:ring_center}(a4) and (a5) show states with two node lines, corresponding to linear combinations of the two degenerate states with winding numbers $m=\pm2$,
$\psi_{+2}^{0}(\vect{r}) \simeq \sqrt{2} u(r)\cos(2\theta-\theta_2)$ and $\psi_{-2}^{0}(\vect{r}) \simeq \sqrt{2} u(r)\sin(2\theta-\theta_2)$, with some angle $\theta_2$.

\begin{figure}[t!]
         \centering
         \includegraphics[width = 0.85\columnwidth]{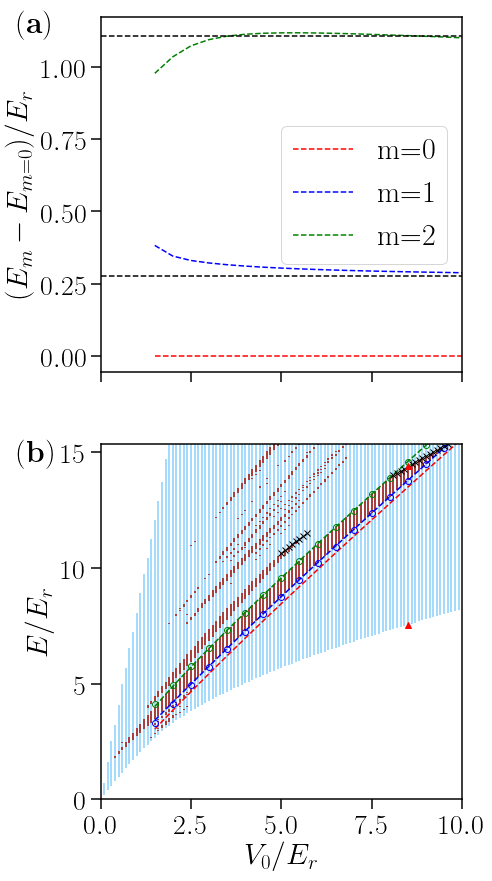}
\caption{\label{fig:gap_compare_n4}
Centered ring states.
(a)~Energy differences $E_m - E_{0}$ between the ring states with winding numbers $m=\pm 1$~(blue) and $\pm 2$~(green), and the ring state with winding number $m=0$~(red, zero by construction).
The dashed black lines show the corresponding theoretical estimates,
${\hbar^2 \, m^2}/{2M\rho_0^2}$, see Eq.~(\ref{eq:ringenergy}), for $m=\pm 1$ and $m=\pm 2$.
(b)~Energy spectrum (reproduced from Fig.~\ref{fig:ratiomin=50}) showing bands (blue) and gaps (red). The dashed lines show the energies of the ring states $E_m$ obtained from diagonalization in a small system with size $L=4a$ for $m=\pm 1$ (blue) and $m=\pm 2$ (green). The circles are the theoretical estimates using Eq.~(\ref{eq:ringenergy}) with $E_0$ corresponding to the numerical value.
The black crosses show the energies of states which enter and finally close the large gaps at large $V_0$, and the two red triangle correspond to the two states plotted in Fig.~\ref{fig:gapclose}, see discussion in Sec.~\ref{sec:gaps.closing}.
}
 \end{figure}

We now check the validity and accuracy of our model.
On the one hand, because of their strong localization, we can find the exact centered ring states by performing diagonalization in a small square around $\rr=0$ with size $L=4a$, larger than the ring diameter $2\rho_0\simeq 1.2a$.
Figure~\ref{fig:gap_compare_n4}(a) shows the eigenenergies hence obtained subtracted by the eigenenergy of the state with winding number $m=0$ (dashed red line, zero by construction), \ie~$E_m - E_{0}$ for $m=\pm 1$ (dashed blue line) and $m=\pm 2$ (dashed green line).
On the other hand, according to our model, the energies of the centered ring states only differ by their orbital rotation energy, which can be written as
\begin{equation}\label{eq:ringenergy}
E_m \simeq E_0 + \frac{\hbar^2 \, m^2}{2M\rho_0^2},
\end{equation}
where we have neglected the $m$-dependence of the radial Hamiltonian as well as the radial extension of $u(r)$ around the radius of the ring, $\rho_0$.
The results of this prediction are shown as dashed black lines in Fig.~\ref{fig:gap_compare_n4}(a), which corresponds to the orbital rotation energy for winding numbers $m=\pm1$ and $m=\pm2$, respectively.
As expected, we find an increasingly better estimate of the ring states energies as the amplitude of the quasicrystal potential increases, owing to a stronger radial confinement.
In addition, the energies of these ring states are also plotted on top of the full spectrum for a large system of size $L=60a$ in Fig.~\ref{fig:gap_compare_n4}(b).
The full spectrum is reproduced from Fig.~\ref{fig:ratiomin=50}, with blue denoting the bands and red the gaps.
The dashed lines show the energies of the ring states $E_m$ with $m=0$~(red), $m=\pm 1$~(blue), and $m=\pm 2$~(green) as obtained from diagonalization in a small system with size  $L=4a$
and the open disks represent the corresponding theoretical estimates, Eq.~(\ref{eq:ringenergy}).
The ring state energies lie on the bands edges.
This is consistent with the fact that the large gap is created by the centered, localized ring states with winding number $m = \pm 1$ and $m= \pm 2$, before it starts to close at $V\simeq 8\Er$.

\subsection{Off-centered rings}\label{sec:gaps.off-centered}
\begin{figure}[t!]
         \centering
         \includegraphics[width = 0.99\columnwidth]{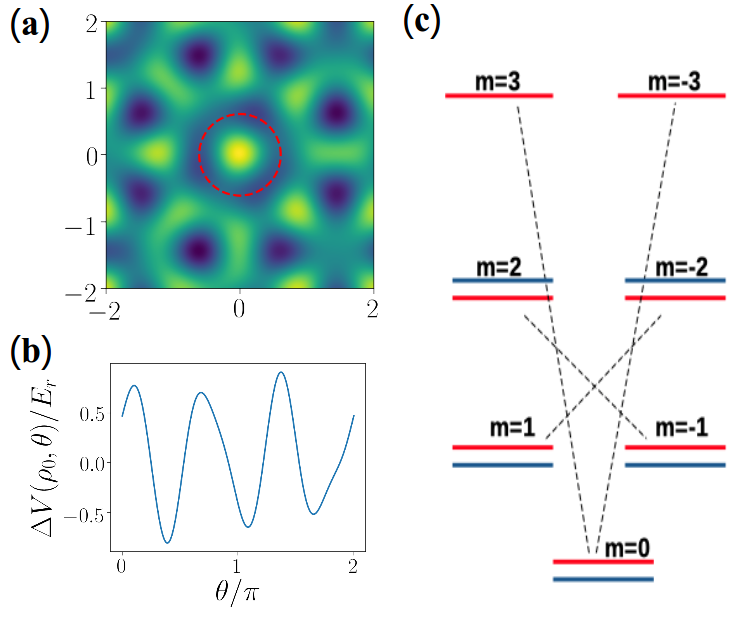}
\caption{\label{fig:repelling}
Off-centered ring states.
(a)~Quasicrystal potential in the vicinity of an off-centered annular potential well, centered at $\rr_0=(19.06a,4.95a)$.
Color scales represent the potential from low values (dark purple) to high values (light yellow).
The dashed circle is the ring of radius $\rho_0 = 0.61a$ around the local maximum.
(b)~Perturbation potential $V(\rho_0,\theta)$ along the ring as a function of the polar angle $\theta$.
(c)~Sketch of the energy-level repulsion picture. The red lines represent energy levels of the centered (unperturbed) ring states with different winding numbers, and the blue lines their off-centered (perturbed) counterparts. The dash lines represent the dominant couplings.
}
 \end{figure}

Let us now examine the off-centered ring states, which are located in annular potential wells around various positions away from the symmetry center of the quasicrystal potential.
A typical example is shown in Fig.~\ref{fig:ring_center}(b),
which corresponds to ring states centered at $\rr_0=(19.06a, 4.95a)$ with $0$, $1$ or $2$ node lines.
Quasiperiodicity of the system implies that the potential around such states is similar to that around the central ring, but with distortions.
The latter are weak but clearly nonnegligible, see Fig.~\ref{fig:repelling}(a).
In particular, the potential and the ring states do not strictly fulfill eight-fold rotation symmetry around the local center at $\rr_0 \neq 0$,
see in particular the appearance of three deeper wells around the central ring (dashed red line).
As a consequence, the corresponding ring states, Figs.~\ref{fig:ring_center}(b1)-(b5),
are not as symmetric as the centered ones, Figs.~\ref{fig:ring_center}(a1)-(a5).
Nevertheless, the off-centered ring states have nearly the same radius as the centred one,  $\rho_0 \simeq 0.61a$, and can still be classified according to their winding number, or equivalently, node line number $m$.
Similar properties are found around other ring states centered at different locations.

As mentioned above, we have observed that the off-centered ring states with winding number $m=0$ or $m=\pm1$ have lower energy than their centered counterparts. In contrast, for winding number $m= \pm 2$, the off-centered ring states have a higher energy than their centered counterparts.
This phenomenon is due to level repulsion, as we show now.
Owing to the similarity of the off-centered annular potential wells and ring states with their centered counterparts, we may understand the properties of the former as a perturbation of the latter.
To lowest pertubation order, we write the off-centered ring states as the shifted centered one,
$\phi_m(\rr) \simeq \phi_m^0(\rr-\rr_0)$,
and the perturbation potential as $\Delta V(\rr) = V(\rr)-V(\rr-\rr_0)$.
Working in the sub-Hilbert space of these states, the perturbation matrix elements are
\begin{equation}\label{eq:perturbation} 
\bra{\phi_{m_1}}\Delta V\ket{\phi_{m_2}} \propto \int d\theta\,  \Delta V(\rho_0,\theta) \e^{i(m_2-m_1)\theta}.
\end{equation}
On the one hand, the left hand side of Eq.~(\ref{eq:perturbation}) can be calculated numerically,
where the required wavefunctions are reconstructed by linear combination of the real-valued numerical wavefunctions, using
$\phi_m(\vect{r})=[\psi_{+m}(\vect{r})+i\psi_{-m}(\vect{r})]/\sqrt{2}$ for $m\in\mathbb{Z}$.
On the other hand, the right hand side is found by neglecting the radial extension of the ring states
and $\Delta V(\rho_0,\theta)$ is a shorthand notation for the perturbed potential along the ring of radius $\rho_0$ with $\theta$ the polar angle.
Due to the three-period oscillation of the potential around the ring shown in Fig.~\ref{fig:repelling}(b),
the Fourier integral in Eq.~(\ref{eq:perturbation}) shows two resonances for $m_2 - m_1 = \pm 3$,
and we expect that the couplings are dominated by the processes such as $m \leftrightarrow m \pm 3$.
We indeed find that the corresponding couplings are at least one order of magnitude larger than the other ones, see perturbation matrix for winding numbers up to $\vert m \vert =3$ in Appendix~\ref{sec:appendix.couplings}.
Since perturbation is stronger for states with closer unperturbed eigenenergies,
the strong couplings between $m \leftrightarrow m \pm 3$ effectively creates
a 2-level system for states $\ket{m=+1}$ and $\ket{m=-2}$ on the one hand,
and for states $\ket{m=-1}$ and $\ket{m=+2}$, on the other hand.
The state $\ket{m=0}$ is strongly coupled to both states $\ket{m=3}$ and $\ket{m=-3}$, hence forming a three-level system.
The overall effective system is sketched in Fig.~\ref{fig:repelling}(c).
Energy level repulsion in the 2-level systems shifts down the energies of the states with winding number $m=\pm1$
and up the energies of the states with winding number $m=\pm2$.
Note that the diagonal perturbation terms, $\bra{\phi_{m}}\Delta V\ket{\phi_{m}}$ are negligible.
The couplings between state $\ket{m=0}$ and states $\ket{m=3}$ and $\ket{m=-3}$ shifts down the energy of $\ket{m=0}$, since off-diagonal perturbations always yields negative corrections to the ground-state energy.

As a result, the energies for the off-centered ring states with windings $m=0,\pm1,\pm2$ are shifted in specific directions, which is in agreement with our numerical diagonalization.
As the largest energy gap corresponds to that between ring states with winding numbers $m = \pm 1$ on the one hand and $m = \pm 2$ on the other hand, we have shown that the bottom of this gap is bounded by the winding $m \pm 1$ ring states at center $\rr=0$.
Likewise, the top of this gap is bounded by the winding $m = \pm 2$ ring states at center $\rr=0$.
This persists in the thermodynamic limit, and the gap width is just the energy difference between these centered ring states.

\subsection{Closing of gaps}\label{sec:gaps.closing}

\begin{figure}[t!]
         \centering
         \includegraphics[width = 0.99\columnwidth]{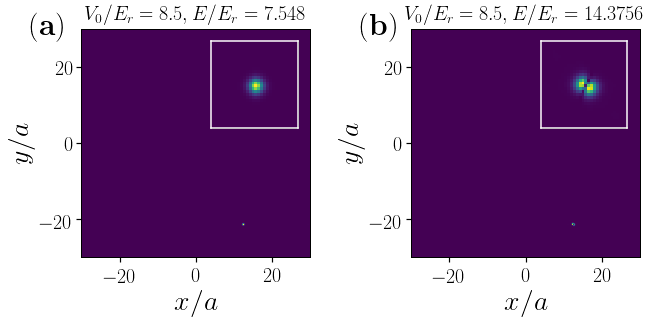}
\caption{\label{fig:gapclose} 
(a)~Ground state of the system alongside
(b)~the first state after the first large energy gap, corresponding to the two red triangles in Fig.~\ref{fig:gap_compare_n4}(b).
The system size is $L=60a$ and the potential amplitude is $V_0=8.5\Er$. The insets show magnifications of the strongly localized states.
}
 \end{figure}

Our model relates the large gaps to ring states with different winding numbers.
According to Eq.~(\ref{eq:ringenergy}), valid for strong radial confinement, the gap widths may thus be expected to reach a constant value for large enough potential depth $V_0$, see also Fig.~\ref{fig:gap_compare_n4}(a).
Consistently, we indeed observe that
the blue and green dashed lines in Fig.~\ref{fig:gap_compare_n4}(b), which show the energies of the centered ring states with winding numbers $m=\pm 1$ and $m=\pm 2$, respectively, are almost parallel to each other.
They match the boundaries of the largest gap, shown in red, for a significant energy range, for $2 \lesssim V_0/\Er \lesssim 8.1$.
However,
this gap, as well as the next one right above, close progressively, respectively in the ranges $8.1\Er \lesssim V_0 \lesssim 9.6\Er$ and $5\Er \lesssim V_0 \lesssim 5.7\Er$,
see the black crosses in Fig.~\ref{fig:gap_compare_n4}(b), which indicate the upper limit of the gaps in these regions.
In fact, the closing of both these gaps is due to another kind of strongly-localized states.
For weak potential amplitude $V_0$, the latter have an energy well inside a energy band above the second gap,
but when $V_0$ increases, they enter successively each gap between the ring states and create the new upper limit of the gaps.
The state closing the largest gap for $V_0=8.5\Er$ [upper red triangle in Fig.~\ref{fig:gap_compare_n4}(b)] is shown in Fig.~\ref{fig:gapclose}(b)
and, for comparison, the ground state of the system (lower red triangle in Fig.~\ref{fig:gap_compare_n4}(b)] is shown in Fig.~\ref{fig:gapclose}(a).
Both these states turn out to be strongly localized in the same deep potential well, and correspond to the ground state and the first excited state of this well.
More precisely, we may approximate the deep potential well by a harmonic potential.
The state in Figure~\ref{fig:gapclose}(a) is then the ground state of the trap, while the state in Figure~\ref{fig:gapclose}(b) is the first excited state of the same trap. 
Consistently with this interpretation, we note that the local minimum of the trap has a vanishingly small potential energy and the energy of the excited state in Fig.~\ref{fig:gapclose}(b) is about twice that of the state in Fig.~\ref{fig:gapclose}(a), see values on top of the figures.
This is what is expected for a 2D isotropic harmonic trap.
The discrepancy to exact energy doubling (about $5\%$) may be attributed to slight anisotropy and/or anharmonicity of the trap.
Such kind of deep potential wells spread over the system and similar states as in Fig.~\ref{fig:gapclose} located around these potential wells are found with similar energies.
As the potential amplitude increases, the first excited states of such deep potential wells, similar to Fig.~\ref{fig:gapclose}(b), enter the higher large gap at about $V_0 \simeq 5\Er$ and comletely close it at about $V_0 \simeq 5.7\Er$. Then, as the potential amplitude further increases, those states enter the lower large gap at about $V_0 \simeq 8.1\Er$ and close it at about $V_0 \simeq 9.6\Er$, see black crosses in Fig.~\ref{fig:gap_compare_n4}(b).

A similar phenomenon explains why we do not observe gaps at a higher energy. In principle, Eq.~(\ref{eq:ringenergy}) suggests even larger gaps at higher energies. However, we find that various kinds of states other than the ring states appear inside the gaps induced by the sole ring states and the latter are not visible.
Moreover, these different kinds of states hybridize at high energy and the structure of the spectrum is not governed by clear ring states any more.

\subsection{Self-similarity and minigaps}\label{sec:gaps.minigaps}
\begin{figure}[t!]
         \centering
         \includegraphics[width = 0.99\columnwidth]{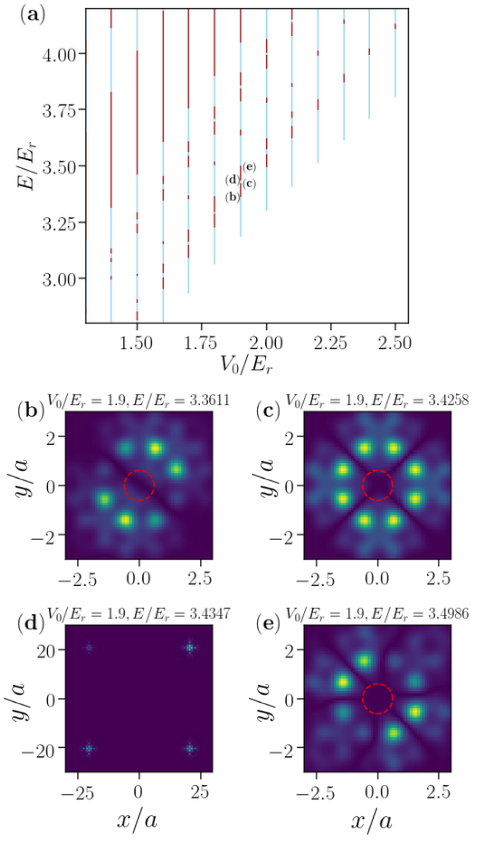}
\caption{\label{fig:gapzoom}
(a)~Zoomed spectrum of the eight-fold quasicrystal potential for various amplitudes $V_0$. Bands are colored blue and gaps are red.
(b)-(e)~Eigenstates at band edges for the two gaps with energy around $E=3.4\Er$ for $V_0=1.9\Er$.
(b),~(c)~Bottom and top states of the first gap.
(d),~(e)~Bottom and top states of the second gap.
}
 \end{figure}

The spectrum in Fig.~\ref{fig:ratiomin=50} presents self-similar structures.
Figure~\ref{fig:gapzoom}(a) shows a zoom of the latter around $V_0=2\Er$ and $E=3.5\Er$.
It clearly shows several gaps, the lowest two for instance around energy $E \simeq 3.4\Er$ for $V_0=1.9\Er$, and different energy for different $V_0$.
Although these gaps are almost invisible on Figs.~\ref{fig:ratiomin=50} and \ref{fig:gap_compare_n4}(b),
we have checked that they are legitimate gaps, according to the procedure presented in Appendix~\ref{sec:appendix.gaps}.

 \begin{figure*}[t!]
    \centering
    \subfigure{\includegraphics[width=0.3\textwidth]{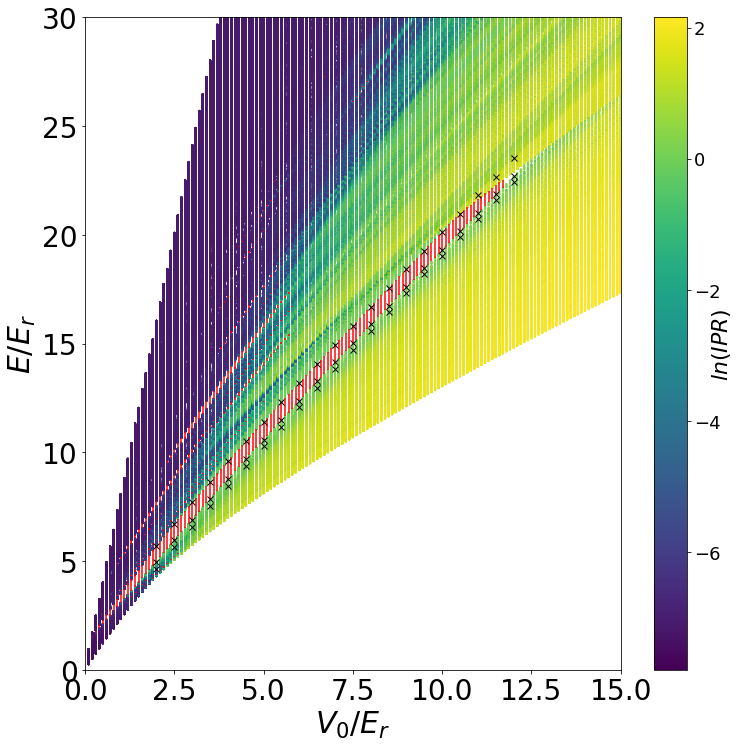}}
~~~    \subfigure{\includegraphics[width=0.30\textwidth]{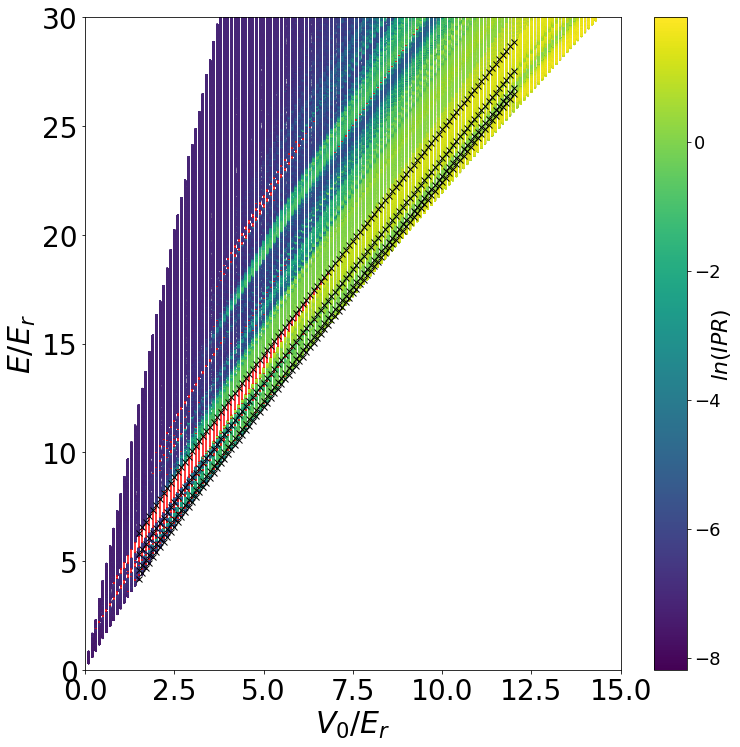}} 
~~~     \subfigure{\includegraphics[width=0.3\textwidth]{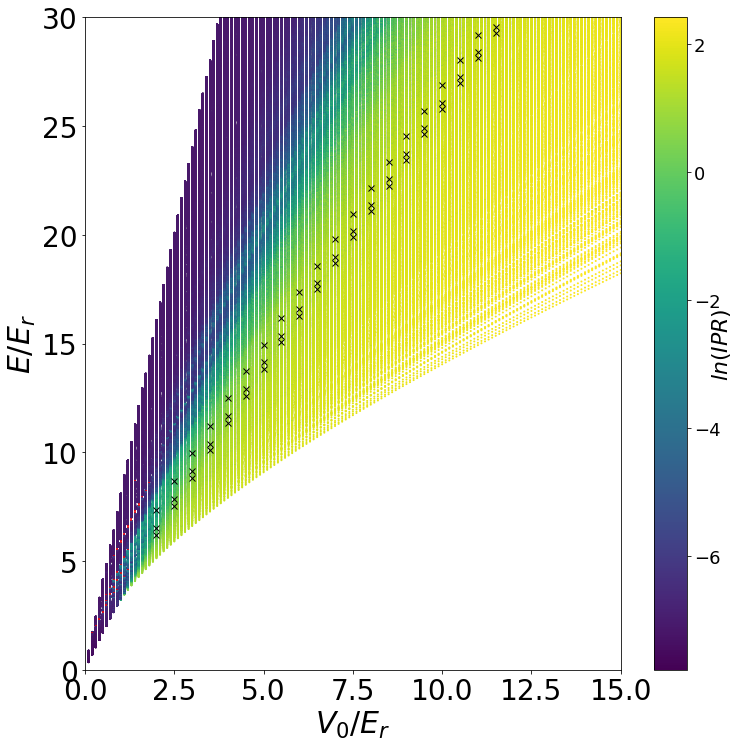}} 
\caption{\label{fig:different_n}
Energy spectra of quasicrystal potentials with different rotational symmetries as created by $n$ pairs of counterpropagating laser beams, with $n=5$ (left, ten-fold), $n=6$ (middlet, welve-fold), and $n=7$ (right, fourteen-fold). The system size is $L=60a$, centered at $x=-13543a, y=196419a$. The color scale represents the value of $\ln(\IPR)$ of each eigenstate, while the gaps are all colored red.
The energies of the first few centered ring states are shown as black crosses.
}
 \end{figure*}

To identity the nature of these gaps, the states at their edges are plotted in Figs.~\ref{fig:gapzoom}(b)-(e). 
Figure~\ref{fig:gapzoom}(b) shows the state at the bottom edge of the first gap. It is a state localized around the quasicrystal center $\rr=0$ and it is composed of eight spots corresponding to eight local potential wells. The latter are identical to each other due to exact eightfold rotational symmetry around $\rr=0$. They lie on a ring larger than the small ring discussed above [shown for reference as a dash red circle in Figs.~\ref{fig:gapzoom}(b), (c), and (e)]. 
The states localized on this larger ring can still be classified according to their winding numbers or equivalently node line numbers.
The state in Fig.~\ref{fig:gapzoom}(b) shows one clear node line and it has thus winding number $m= \pm 1$. 
Similarly Fig.~\ref{fig:gapzoom}(c) shows the state with winding number $m= \pm 2$, which lies near the top of the first gap,
and the state at the top of the second gap shown in Fig.~\ref{fig:gapzoom}(e) that with winding number $m= \pm 3$.
Note that the state of Fig.~\ref{fig:gapzoom}(c) is not strictly at the top of the first gap. The state at the very top edge of the first gap is composed of four off-centered copies of Fig.~\ref{fig:gapzoom}(c) connected together and has a slightly lower energy than the centered state shown in Fig.~\ref{fig:gapzoom}(c). The state at the bottom edge of the second gap shown in Fig.~\ref{fig:gapzoom}(d) is not localized at $\rr=0$ and has no clear special structure.

The general picture of these small gaps is thus similar to that of the large gaps discussed above created by the small ring states with different winding numbers. The only difference is that the ring states creating the small gaps have a larger ring radius, so that the energy differences due to phase windings are much smaller, and the gap sizes are comparatively much smaller than the gaps created by the small rings.

\section{Quasicrystal potentials with different rotational symmetries}\label{sec:generalization}
So far, we have considered the localization and spectral properties of a quasicrystal potential with eightfold rotational symmetry.
Here we briefly discuss whether or not quasicrystals with different rotational symmetries may also possess similar properties.
To answer this, we study quasicrystal potentials that are generated by interfering a larger number $n$ of laser beams.
The general formula of a quasicrystal potential with a $2n$-fold rotational symmetry is
\begin{equation}\label{eq:QPpotential_n}
V(\rr) = V_0 \sum_{k=1}^n \cos^2 \left(\vect{G}_k \cdot \rr +\phi_k \right),
\end{equation}
with $\vect{G}_k=\frac{\pi}{a}\big(\cos(k{\pi}/{n}) \, ,\sin(k{\pi}/{n})\big)$.
Realization of such potentials have been recently proposed in Ref.~\cite{kosch2022}.
In Fig.~\ref{fig:different_n}, we plot the energy spectra for $n=5$, $6$, and $7$.
The general localization picture is similar to that previously discussed for the eightfold quasicrystal potential ($n=4$):
For sufficiently large amplitude $V_0$, the eigenstates are typically localized at low energy and extended at high energy.
At intermediate energy, we find a nonmonotonous energy-dependence of the localization strength.
Moreover, for $n=5$ or $n=6$, the spectrum also has at least one large gap, similar to that found for $n=4$, as well as smaller but visible gaps.
For $n=7$ instead, the gaps become much narrower and almost invisible at the scale of Fig.~\ref{fig:different_n}.
To understand the origin of these gaps, we note that the quasicrystal potentials created by more than 4 laser pairs also have an annular potential well around the center at $\rr=0$, with almost the same radius, $\rho_0 \simeq 0.61a$.
As a result, this annular potential well also hosts ring states with different winding numbers, the energies of which are still approximately given by Eq.~(\ref{eq:ringenergy}).
The energy differences are thus almost the same as for $n=4$, but the reference energy $E_0$ may be dependent on $n$.
Moreover, they are almost independent of the potential amplitude for large enough $V_0$.
The energies of the first few centered ring states for each quasicrystal potential are plotted as black crosses in Fig.~\ref{fig:different_n}.

For $n=5$, the structure of the spectrum is very similar to that discussed above for $n=4$. In particular, the largest gap is also created by the gap between the ring states with winding numbers $m = \pm 1$ and $m = \pm 2$, and the gap width is almost the same as for $n=4$.
For $n=6$, the largest gap may also be related to ring states, now with winding numbers $m=\pm 2$ and $m=\pm 3$, and the gap size is larger.
More precisely, the top of the gap is indeed composed of centered ring states with winding number $m= \pm 3$.
In contrast, the bottom of the gap is not strictly a centered ring state with winding number $m= \pm 2$, since off-centered ring states with the same winding number have higher energies than the centered one, as discussed above.
For a limited range of quasicrystal amplitude, $2\Er \lesssim V_0 \lesssim 4\Er$, the centered and off-centered ring states with $m=\pm 2$ lie near by the band edge, but the true edge state turns out to be a different state, which can be either localized or extended with a complex structure, depending on $V_0$.
For $V_0 \gtrsim 4\Er$, new localized states enter the bottom of the gap, and progressively closes it from below.
Note also that
for a certain range of $V_0$, roughly between $2\Er$ and $7\Er$, the ground state of the whole spectrum is the centered ring state with winding number $m=0$.
Finally, we find that for $n=7$ the gaps have almost negligible sizes, and there is no large gap near the energies of the ring states.
This is because other kinds of states coexist with those ring states in the same eigenenergy ranges, even though ring states still have the same energy differences for different winding numbers.
For larger $n$, we found that large gaps generated by ring states remain closed.

\section{Conclusion and discussion}\label{sec:conclusions}
In summary, we have shown that 2D optical quasicrystals host exotic localization properties and intriguing spectral features. For the eight-fold rotationally symmetric potential, a finite-size scaling analysis of the $\IPR$ reveals the presence of localized, critical, and extended states.
Extended states dominate the energy spectrum at large energies, while localized states populate the low energy spectrum, as expected.
However, at intermediate energies, we find that localized states can appear alongside critical states, with no clear separation between them. Furthermore, large gaps appear in the spectrum, with some states at the band edges having a strongly localized profile.
These localized states can take the form of ring states with different quantized winding numbers. By modelling these ring states, we have found that the band edges coincide with the theoretical energy of ring states, thus confirming that they play an important role in the formation of energy gaps within optical quasicrystals. Finally, we have also confirmed that quasicrystals with other rotational symmetries can also possess similar kinds of localization properties and energy spectra, with ring states again playing an important role. In all cases, the most prominent gaps of the spectra are stable across a range of potential depths $V_0$ and rotational symmetries, provided that other localized states do not compete or enter the gap generated by the ring states.
Our results shed new light on localization and spectral properties of optical quasicrystal potentials, as realized in recent experiments~\cite{viebahn2019,sbroscia2020,JrChiunYu2023,kosch2022}.
Further application and development of this work may be expected in two directions.

On the one hand, our results are directly applicable to the above-mentioned experiments.
The eightfold quasicrystal potential studied in the main part of the paper has been implemented in the experiments reported in Refs.~\cite{viebahn2019,sbroscia2020,JrChiunYu2023} and potentials with higher order discrete rotation symmetry can be implemented in a similar manner by using a suitable number of laser beams, see for instance Refs.~\cite{lsp2005,kosch2022}. The quasicrystal potential amplitudes considered here, $V_0 \simeq 1-15\Er$, are also relevant for these experiments and inter-atomic interactions can be cancelled with high accuracy using Feshbach resonance methods~\cite{chin2010,pollack2009}.
Localization may be unveiled in ultracold atomic gases using expansion schemes, as proposed in Refs.~\cite{lsp2005,lsp2007} and realized for instance in Refs.~\cite{billy2008,roati2008,kondov2011,jendrzejewski2012}, see also Refs.~\cite{kuhn2007,skipetrov2008,piraud2011,piraud2012a,piraud2013b} for further theoretical discussions.
In this scheme, an initially trapped ultracold-atom gas is released into the quasicrystal potential, generating a wavepacket covering a tunable range of energy components. The components whose energy corresponds to a band of localized states stop expanding on a short length scale, while those whose energy corresponds to a band of extended states show normal diffusive expansion. Direct imaging at different times can thus be used to distinguish between them. For the 2D quasicrystal lattice considered here, bands of critical states also exist, for which we can anticipate anomalous diffusion, also observable in the expansion dynamics.
In such scheme, a high-energy cut-off is set by the chemical potential of an initially interacting Bose-Einstein condensate or the Fermi energy of an ultracold gas of fermions.
When controlled by the initial interaction strength and/or the number of atoms, we expect to observe a localized gas at low chemical potential, anomalous diffusion on top of a localized component at intermediate chemical potential, and an additional normal diffusion when the spectral range contains extended states. Note, however, the coexistence of localized, critical, and extended components may make their segregation difficult.
To overcome this issue,
fine selection of a particular energy may be realized using radio-frequency transfer from an internal atomic state insensitive to the quasicrystal potential towards a sensitive state~\cite{pezze2011a}.
Here, the width of the selected energy range is proportional to the inverse of the pulse duration and can be chosen from a band of either localized, critical or extended states. Expansion then leads, respectively, to pure localization, anomalous diffusion or normal diffusion. The non-transferred part undergoes ballistic expansion or can even be eliminated by transfer to a non-imaged internal atomic state. This makes it possible to reconstruct bands of localized, critical, and extended states, as shown in Fig.~\ref{fig:scaling_full}.
The existence of the gaps discussed here can also be demonstrated by this approach.

On the other hand, our results on localization and spectral properties of the single-particle problem studied here also play an important role in the physics of correlated quantum gases in a 2D quasicrystalline potential.
This is particularly the case for a gas of correlated bosons.
In one dimension and in the regime of strong interactions, a gas of bosons can be exactly mapped onto a gas of free fermions, a phenomenon known as fermionization. This makes it possible to map Mott insulators onto free spectral gaps and Bose glass onto localized states.
In dimensions higher than one such exact mapping breaks down, but fermionization persists nonetheless when the Bose gas populates only states that are spatially separated from one another. In this case, strong repulsive interactions suppress multiple occupation of each localized state, hence mimicking an effective Pauli principle in real space.
Our results could thus help understand the onset of a Bose glass as well as a Mott plateau in the strongly-interacting regime, found in recent work~\cite{zhu2023}.

\begin{acknowledgments}
We thank Hepeng Yao for fruitful discussions.
We acknowledge the CPHT computer team for valuable support.
This research was supported by
the Agence Nationale de la Recherche (ANR, project ANR-CMAQ-002 France~2030),
the program ``Investissements d'Avenir'' LabEx PALM (project ANR-10-LABX-0039-PALM),
the IPParis Doctoral School, and HPC/AI resources from GENCI-TGCC (Grant 2022-A0110510300).
\end{acknowledgments}

\appendix

\section{Density of state and gaps}\label{sec:appendix.gaps}
 \begin{figure}[t!]
         \centering
         \includegraphics[width = 0.99\columnwidth]{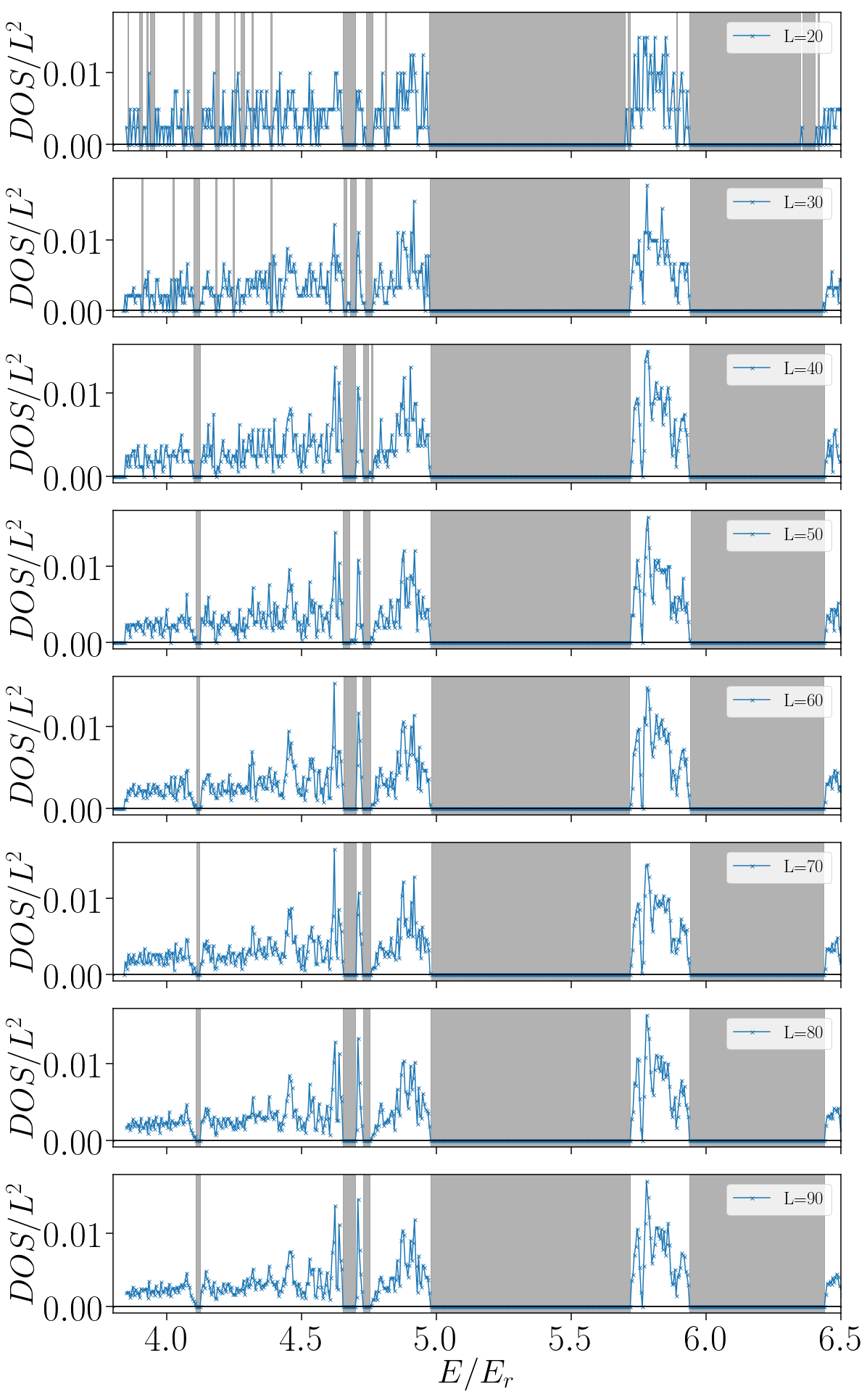}
\caption{\label{fig:spectrum_sizes} 
Density of states per unit area versus energy for the quasicrystal amplitude $V_0=2.5\Er$ and the energy resolution $\epsilon=0.005\Er$.
The various panels correspond to increasing system size from $L=20a$ (top) to $L=90a$ (bottom).
The energy gaps are indicated by grey shaded areas.
}
 \end{figure}
Careful identification of the true energy gaps in the spectrum can be captured by performing a finite-size scaling analysis.
Here we compute the density of state (DOS) using
\begin{equation}
g_\epsilon(E) = \frac{\delta W(E,\epsilon)}{\epsilon},
\end{equation}
where $\epsilon=0.005\Er$  is a finite energy resolution and $\delta W(E,\epsilon)$ is the number of states in the energy window $[E,E+\epsilon)$.
A good estimate of the DOS is obtained when the energy resolution $\epsilon$ is smaller than the typical variation scale of the DOS and larger than the inverse of the typical DOS so that several states are in each energy slice, $g_\epsilon(E) \epsilon \gg 1$.
For the homogeneous 2D gas, we have $g(E)^{-1}\Er=4a^2/\pi L^2 \simeq 0.003$ for the smallest system size $L=20a$. 

Figure~\ref{fig:spectrum_sizes} shows the DOS per unit area versus energy for different system sizes from $L=20a$ to $90a$.
Each point with zero DOS gives an energy gap, so that all gaps larger than the energy resolution $\epsilon$ are revealed.
For the smallest system size of $L=20a$ in Fig.~\ref{fig:spectrum_sizes}, the DOS displays many small gaps and a few large energy gaps. While the two large gaps are stable against increasing the system size, only a few small gaps survive for large system sizes.
For the energy resolution $\epsilon=0.005\Er$, the structure of the spectrum is stable when the system size is larger than $L \simeq 60a$.

\section{Linear fit quality for finite-size scaling of the IPR}\label{Appendix:QualityFitIPR}
 \begin{figure}[t!]
         \centering
         \includegraphics[width = 0.99\columnwidth]{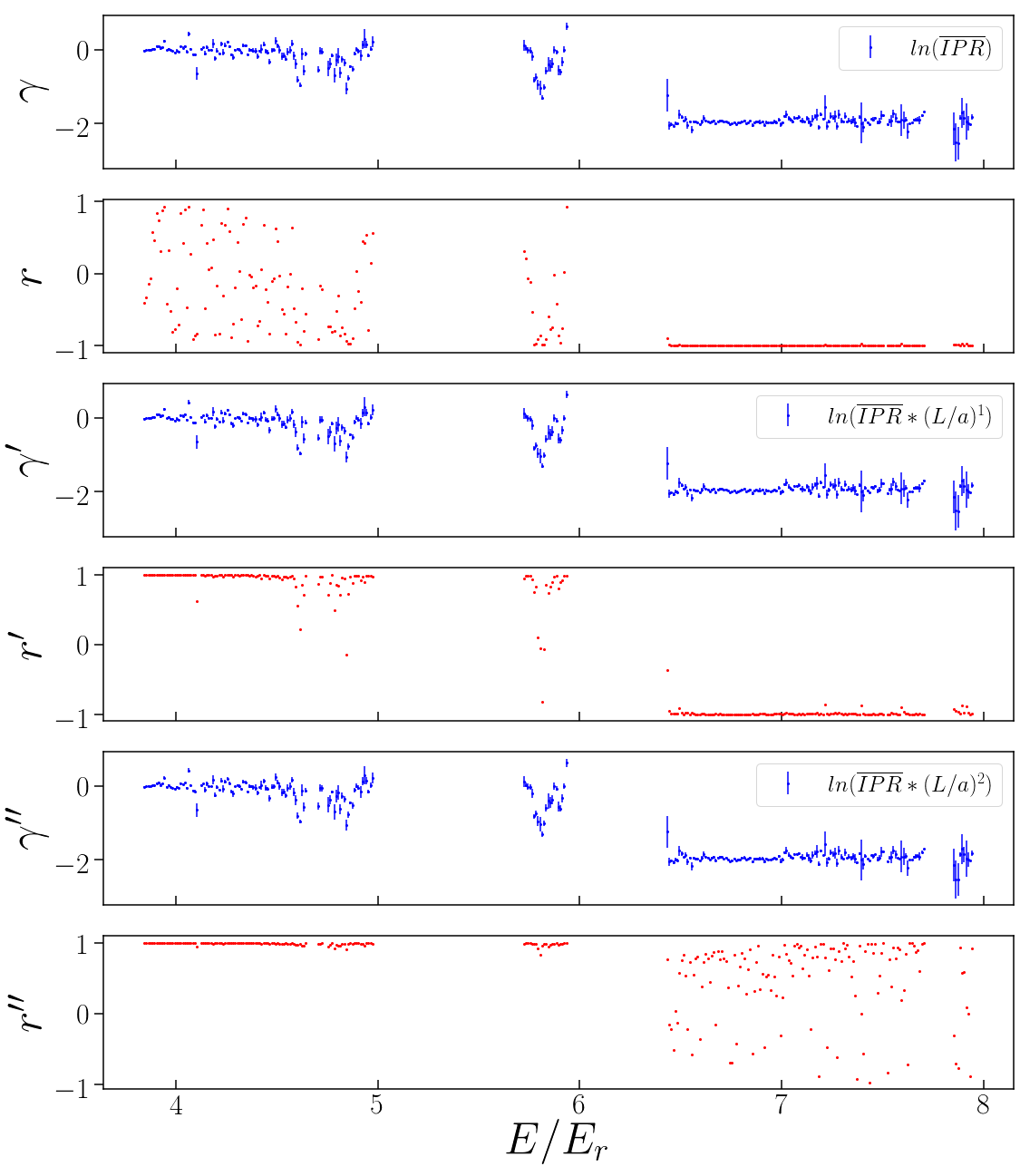}
\caption{\label{fig:scaling_fullspectrum_appendix} 
Finite size scaling of the $\IPR$, performed as in Fig.~\ref{fig:scaling_single},
for system sizes ranging from $L=20a$ to $L=100a$, except in some narrow energy windows where data for larger sizes, $L=120a,160a,200a,300a$, are also calculated.
The panels, from top to bottom, show, successively the scaling exponent and the corresponding regression coefficient:
(i)~$\gamma$ and $r$ for $\ln(\overline{\IPR})$ versus $\ln(L/a)$,
(ii)~$\gamma^\prime$ and $r^\prime$ for $\ln(\overline{\IPR}\cdot L/a)$ versus $\ln(L/a)$,
and (iii)~$\gamma^{\prime\prime}$ and $r^{\prime\prime}$ for $\ln(\overline{\IPR}\cdot (L/a)^2)$ versus $\ln(L/a)$.
}
 \end{figure}
The exponent $\gamma$, such that $\IPR \sim L^\gamma$, is found from linear fits of $\ln(\overline{\IPR})$ versus $\ln(L/a)$, with results shown in the first row of Fig.~\ref{fig:scaling_fullspectrum_appendix} (same data as in the upper low of Fig.~\ref{fig:scaling_full} in the main text).
To characterize the fit quality, we compute the Pearson correlation coefficient,
\begin{equation}
    r=\frac{\sum_j( x_i-\overline{x} )(y_i-\overline{y})}{\sqrt{\sum_i ( x_i-\overline{x} )^2 \sum_i (y_i-\overline{y})^2}},
\end{equation}
for two data sets $x_i$, $y_i$ where $\overline{x}$ and $\overline{y}$ are their mean values.
A value of $\lvert r \rvert$ close to $1$ indicates good linear correlation between the data sets, while $\lvert r \rvert$ close to $0$ indicates poor linear behaviour. The correlation coefficient for linear fitting of $\ln(\overline{\IPR})$ and $\ln(L/a)$ is shown in the second row of Fig.~\ref{fig:scaling_fullspectrum_appendix}. It shows good linear behaviour with $r\simeq-1$ and $\gamma\simeq-2$ for energy $E\gtrsim 6.45\Er$, and the states in this energy range are clearly extended.
For a lower energy range, however, the coefficients $r$ are much worse, which questions the corresponding results.
This is actually misleading as when the slope $\gamma$ is about $0$, even weak fluctuations of data points around a constant value can greatly affect the coefficients $r$.

 \begin{figure}[t!]
	\centering
	\includegraphics[width = 0.99\columnwidth]{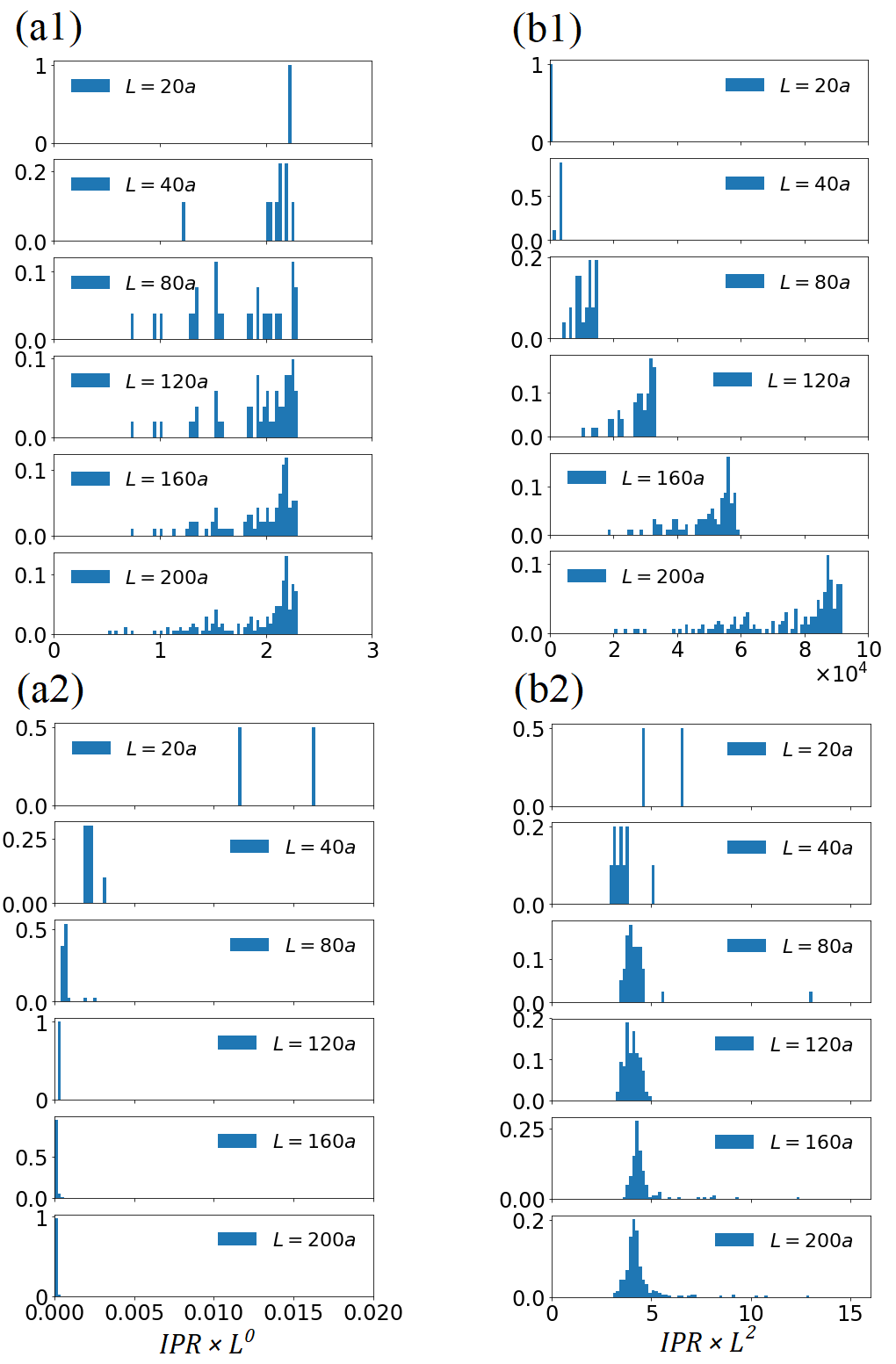}
	\caption{\label{fig:ipr_spread_le} 
		Distributions of $\IPR$ for different system sizes $L$, where the horizontal axis is rescaled as $\IPR \times L^{-\gamma}$,
		with $\gamma=0$~(left column) and $\gamma=-2$~(right column), for energy windows (upper row)~$E/\Er \in [4.00, 4.01]$ in the localized domain and (lower row)~$E/\Er \in [6.45,6.46]$ in the extended domain.
	}
\end{figure}

To circumvent this issue, we repeat the linear regression for $\ln(\overline{\IPR}\cdot L/a)$ versus $\ln(L/a)$ with slope $\gamma^\prime+1$ and for $\ln(\overline{\IPR}\cdot (L/a)^2)$ versus $\ln(L/a)$ with slope $\gamma^{\prime\prime}+2$.
The results, together with the corresponding correlation coefficients, $r^\prime$ and $r^{\prime\prime}$ respectively, are shown in the lower rows of Fig.~\ref{fig:scaling_fullspectrum_appendix}.
The three linear regressions give almost indistinguishable results for the exponent $\gamma$, $\gamma^\prime$, and $\gamma^{\prime\prime}$, and, for each energy $E$, at least one of the three correlation coefficients $r$, $r^\prime$ or $r^{\prime\prime}$ is close to $1$. 
For instance, the coefficient $\lvert r^{\prime\prime} \rvert$ has values away from $1$ for $E\gtrsim 6.45\Er$, since the slope $\gamma^{\prime\prime}+2$ for $\ln(\overline{\IPR}\cdot (L/a)^2)$ is about $0$. It is, however, close to $1$ for low energy states, and the scaling given by $\gamma^{\prime\prime}$ (essentially equal to $\gamma$ and $\gamma^\prime$) is reliable in this regime.
In general, the fitted slope $\gamma$ yields the correct scaling and the quality is assessed by either $r$, $r^\prime$ or $r^{\prime\prime}$.

So far, we have not discussed the spread of IPR values in a given energy window. For extended and localized domains, we find that this will not greatly influence our results. To show this, we plot in Fig.~\ref{fig:ipr_spread_le} histograms of IPR occurrences for different system sizes, for the same energy windows as in Fig.~\ref{fig:scaling_single} for a localized state (upper row) and an extended state (lower row).
The horizontal axis is rescaled as $\IPR \times L^{-\gamma}$ with $\gamma=0$ (left column) or $\gamma=-2$ (right column).
We find that the distributions are peaked, and remain fixed in position as a function of $L$ for the correct choice of scaling ($\gamma=0$ for the localized state and $\gamma=-2$ for the extended state). In contrast, for the opposite choice of the scaling, the position drifts.
This immediately verifies that the underlying states are either localized or extended.

 \begin{figure}[t!]
	\centering
	\includegraphics[width = 0.99\columnwidth]{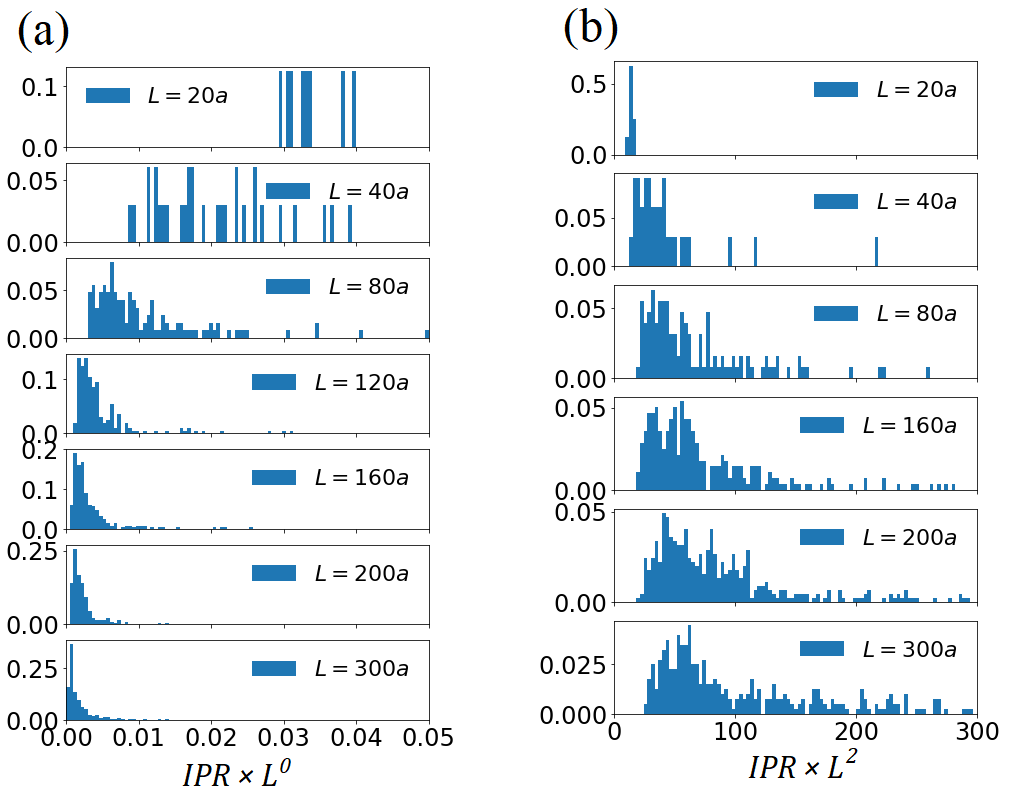}
	\caption{\label{fig:ipr_spread_c}
		Distributions of $\IPR$ for different system sizes $L$, where the horizontal axis is rescaled as $\IPR \times L^{-\gamma}$,
		with $\gamma=0$~(left) and $\gamma=-2$~(right), for energy windows $E/\Er \in [5.81, 5.82]$ in the critical domain.
	}
\end{figure}

The behaviour of critical states, however, is more complicated. In Fig.~\ref{fig:ipr_spread_c}, we consider the critical energy window of Fig.~\ref{fig:scaling_single}. The spread of IPR values is much larger, which would of course inflate the relative errorbars of the average IPR.
However, we observe that the distribution clearly shows a single peak. Moreover, the 
distributions are shifted for both rescalings $\gamma=0$ or $\gamma=-2$ for increasing $L$. This indicates that a majority of states in the considered energy window are neither localized or extended, but critical.

\section{Structures within critical states}\label{sec:appendix.critical_square}
The two kinds of critical states shown in Fig.~\ref{fig:wavefun_critical1} and Fig.~\ref{fig:wavefun_critical2}
both have the ring states discussed in Sec.~\ref{sec:gaps}
as the building blocks. The ring-shaped critical states shown in Fig.~\ref{fig:wavefun_critical1} have the ring states with winding $m=0$ as the building blocks, see Figs.~\ref{fig:ring_center} (a1) and (b1), note that these critical states have energy slightly smaller than Figs.~\ref{fig:ring_center} (a1) and (b1). The building blocks of square-shaped critical states shown in Fig.~\ref{fig:wavefun_critical2} are the ring states with winding number $m=2$, see Figs.~\ref{fig:ring_center} (a4), (b4), (a5), and (b5), and these critical states have energy slightly larger than Figs.~\ref{fig:ring_center} (a4), (b4), (a5), and (b5). As we discussed in Sec.~\ref{sec:gaps}, the off-centered ring states have lower energy than the centered one for $m=0$ and higher energy than the centered one for $m=2$. At a certain point, off-centered ring states at different locations hybridize together, and forming the critical states. 

\begin{figure}[t!]
         \centering
         \includegraphics[width = 0.99\columnwidth]{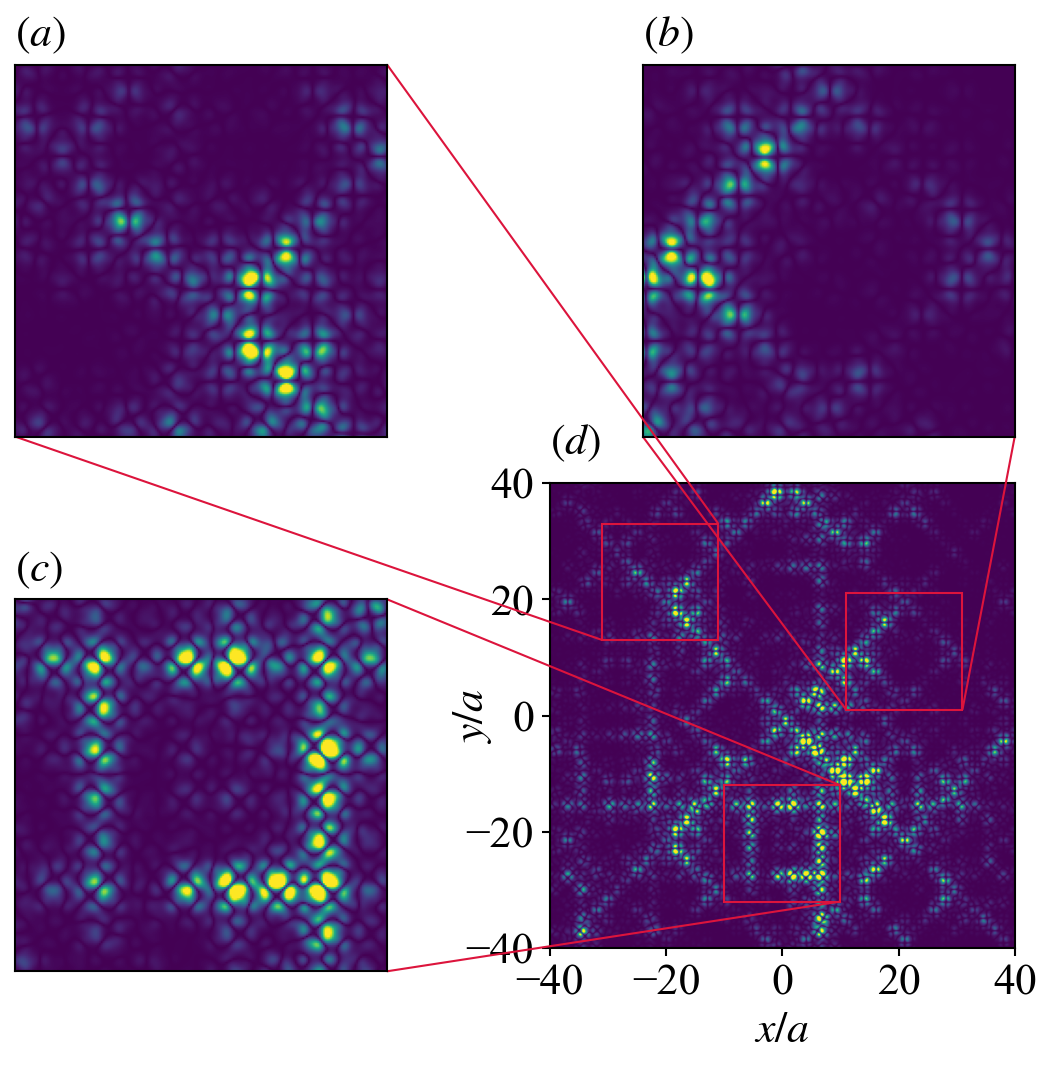}

\caption{\label{fig:critical_zoom_appendix} 
(a), (b), and (c): Zoomed up plots of several regions of the wavefunction shown in (d), which is a copy of Fig.~\ref{fig:wavefun_critical2}(d).
}
 \end{figure}

The square shape of the critical states in Fig.~\ref{fig:wavefun_critical2} can be understood from the shape of the $m=2$ ring states, as shown in Figs.~\ref{fig:ring_center} (a4), (b4), (a5), and (b5). They may have node lines in diagonal directions as Figs.~\ref{fig:ring_center} (a4) and (b4) or in horizontal and vertical directions as Figs.~\ref{fig:ring_center} (a5) and (b5). When these $m=2$ ring states couple to each other, they tend to connect in the direction where the bright spots touch, as this will give a strong coupling. As a result, $m=2$ ring states with node lines in diagonal directions couple to each other along horizontal and vertical directions, and ring states with node lines in horizontal and vertical directions couple to each other along diagonal directions. Fig.~\ref{fig:critical_zoom_appendix} shows several regions of the wavefunction Fig.~\ref{fig:wavefun_critical2} (d), plotted in zoomed up scales. We can see that the typical directions of the state structure are indeed different from the directions of the node lines in each of the unit spots.

 \begin{figure}[t!]
	\centering
	\includegraphics[width = 0.99\columnwidth]{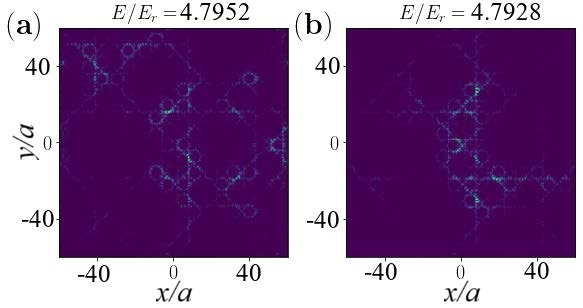}
	\caption{\label{fig:wavefun_critical2b} 
		Examples of critical states with energy near $E \simeq 4.8\Er$, which are composed of $m=1$ ring state components (i.e. a single node line). We consider $V_0=2.5\Er$ and $L/a=100$. Localized components become coupled in circular, ring-like arrangements. These circular structures are then coupled across diagonal lines.
	}
\end{figure}

Finally, in Fig.~\ref{fig:wavefun_critical2b}, we show two examples of critical states with energy near $E\simeq4.8\Er$. These states are composed of hybridized ring-like structures on small scale, with a similar square-shaped pattern to what has been observed before on large scale. At the intermediate length scale, we observe circular structures composed of several ring states with a single node line each (\ie~$m=1$ ring states). The node lines are aligned in such a way to face the centre of the circle. Different circles couple together and form extensive critical states.


\section{Ring states and winding number}\label{app:Ringstates}
We discuss here the main properties of the ring states.

\subsection{Centered ring states}\label{sec:appendix.centeredring}
The centered ring states live in an annular potential valley centered on $\rr=0$ with radius $\rho_0$.
In the vicinity of the potential valley, the angular dependence of the potential is of the order of $0.025V_0$, much smaller than any typical energy scale in this system, leading to wavefunction modulations less than $10\%$ for $V_0 \lesssim 10\Er$. We can thus neglect it and write $V(\rho, \theta) \simeq V(\rho)$.
Owing to rotation symmetry, the local Schr\"odinger equation and the planar angular momentum operator, $\hat{L}_z=-i\hbar \frac{\partial}{\partial \theta}$, may be diagonalized simultaneously.
The energy eigenstates read as $\phi(\rho,\theta)=u_m(\rho)\e^{i m\theta}$, with $m \in\mathbb{Z}$.
Writing $u_m(\rho) = \rho^{-1/2} f_m(\rho)$, the amplitude $f_m(\rho)$ is then governed by the semi-infinite one-dimensional, radial equation
\begin{equation}
-\frac{\hbar^2}{2M} \frac{\dd^2}{\dd \rho^2}f_m(r) + V_m(\rho) f_m(\rho) = E_m f_m(\rho), \quad \rho > 0,
\end{equation}
with the effective potential
\begin{equation}
V_m(\rho) = V(\rho) + \frac{\hbar^2}{2M} \frac{4m^2-1}{4\rho^2}.
\end{equation}
The latter consists of the bare potential $V(\rho)$ and a centrifugal term.

\begin{figure}[t!]
\centering
\includegraphics[width=0.9\columnwidth]{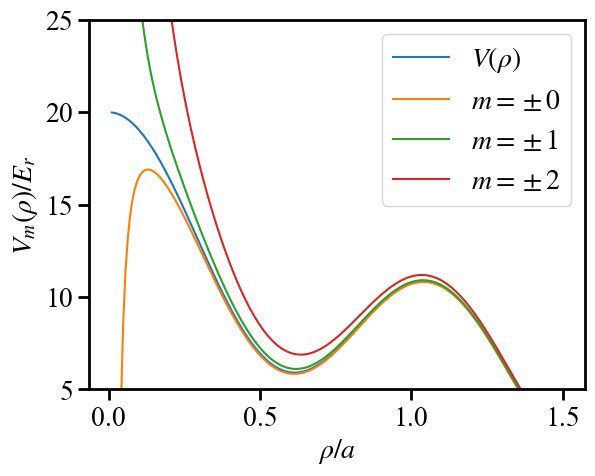}
\caption{Effective potential $V_m(\rho)$ for a quasicrystal potential with amplitude $V_0=5\Er$ and for varius winding numbers $m$.
}
\label{fig:Veff}
\end{figure}

\begin{figure}[t!]
\centering
\includegraphics[width=0.9\columnwidth]{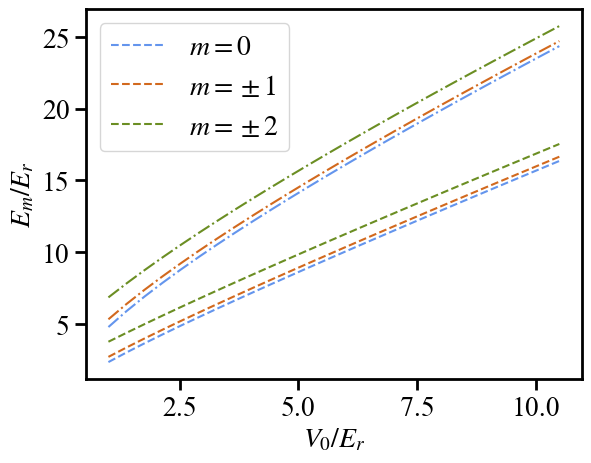}
\caption{\label{fig:Em}
Eigenenergies of the centered ring states in the radial ground state ($n=0$, dashed lines) and first excited state ($n=1$, dotted-dashed lines) with winding numbers $m=0$ (blue), $\pm 1$ (red), and $\pm 2$ (green).
}
\end{figure}

As shown in Fig.~\ref{fig:Veff}, the centrifugal term strongly deforms the potential for $\rho \lesssim 0.1a$, but the distortion near the minimum, $\rho_0\simeq 0.61a$, is weak enough that the radial eigenfunction $f_m(\rho)$ weakly depends on $m$ for sufficiently large $V_0$ and small $m$.
For a sufficiently deep potential well, we may use a harmonic approximation and write
\begin{equation}
V_m (\rho) \simeq V_m (\rho_0) + \frac{1}{2}M\omega_m^2 (\rho - \rho_0)^2.
\end{equation}
The dependence of $\omega_m$ on $m$ is less
than $10\%$ for $V_0>3\Er$ and $\vert m \vert \leq 2$, and we can omit the subscript.
Similarly, the radius of the local minimum is almost independent of $m$ in the same range of parameters.
The extension of the radial ground-state wavefunction is given by $\Delta\rho = \sqrt{\langle (\rho-\rho_0)^2 \rangle} \simeq \sqrt{\hbar / 2M\omega}$.
For $V_0 \gtrsim 3\Er$, it is small enough, $\Delta\rho \lesssim 0.2a$, that the harmonic approximation is good.
As a result, the low-energy centered ring eigenstates have energies
\begin{equation}
E_{m,n} \simeq V(\rho_0) + \hbar \omega(n+1/2) + \frac{\hbar^2}{2M}\frac{4m^2-1}{4\rho_0^2},
\end{equation}
where $n \in\mathbb{N}$ denotes the $n$-th radial excited state.
These eigenenergies are plotted in Fig.~\ref{fig:Em} for $n=0$ and $1$, and $\vert m \vert \leq 2$.
In the range of interest, we find that the lowest energy states are, successively, $m=0$, $m=\pm 1$, and $m=\pm 2$, all with $n=0$.
Hence, the radial excitations have a higher energy, and we may restrict to the $n=0$ sector.
The relevant ring state only differ by their orbital rotation energy, getting the theoretical estimate
\begin{equation}
E_m - E_0 = \frac{\hbar^2}{2M} \frac{m^2}{\rho_0^2},
\end{equation}
with $E_0 \simeq V(\rho_0) + \hbar\omega/2-\hbar^2/8M\rho_0^2$.
This is nothing but Eq.~(\ref{eq:ringenergy}).

\subsection{Perturbative coupling terms for off-centered ring states}\label{sec:appendix.couplings}
\begin{figure}[t!]
         \centering
         \includegraphics[width = 0.9\columnwidth]{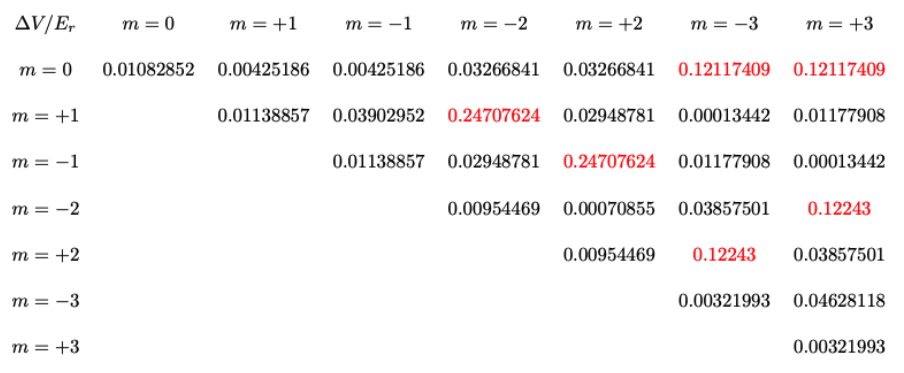}
\caption{\label{fig:matrix} 
Perturbation matrix of the ring states.
}
 \end{figure}

Figure~\ref{fig:matrix} shows, in units of $\Er$, the moduli of all the perturbation matrix elements $\bra{\phi_{m_1}}\Delta V\ket{\phi_{m_2}}$, see Eq.~(\ref{eq:perturbation}), between the 7 ring states with winding numbers $m=0$, $\pm1$, $\pm2$, and $\pm3$. The perturbation $\Delta V$ is the difference between the potentials around $\rr_0=(19.06a, 4.95a)$ and around $\rr=0$.
Since the matrix is symmetric, only the upper half is shown.
The numbers in red are the strongest couplings while those in black are at least about one order of magnitude smaller.

\bibliographystyle{revtexlsp}
\bibliography{biblioLSP,notes}

\end{document}